\documentclass[10pt]{article}
\usepackage{amsmath}
\usepackage{latexsym}
\usepackage{amssymb}
\usepackage{amsfonts}
\usepackage{amsthm}
\usepackage{graphicx}
\title{FROM ELASTICITY TO ELECTROMAGNETISM : BEYOND THE MIRROR}
\author{J.-F. Pommaret \\ CERMICS, Ecole des Ponts ParisTech, France \\
 jean-francois.pommaret@wanadoo.fr  \\
( http://cermics.enpc.fr/$\sim$pommaret/home.html )}
\date{  }
\begin{document}
\maketitle

\thispagestyle{empty}

\noindent
{\bf ABSTRACT}\\

The first purpose of this short but striking paper is to revisit Elasticity (EL) and Electromagnetism (EM) by comparing the structure of these two theories and examining with details their well known couplings, in particular {\it piezoelectricity} and {\it photoelasticity}. Despite the strange Helmholtz and Mach-Lippmann analogies existing between them, no classical technique may provide a common setting. However, unexpected arguments discovered independently by the brothers E. and F. Cosserat in 1909 for EL and by H. Weyl in 1918 for EM are leading to construct a new differential sequence called {\it Spencer sequence} in the framework of the formal theory of Lie pseudogroups and to introduce it for the {\it conformal group} of space-time with 15 parameters. Then, all the previous explicit couplings can be deduced abstractly and one must just go to a laboratory in order to know about the coupling constants on which they are depending, like in the Hooke or Minkowski constitutive relations existing respectively in EL or EM. We finally provide a new combined experimental and theoretical proof of the fact that any $1$-form with value in the second order jets ({\it elations}) of the conformal group of space-time can be uniquely decomposed into the direct sum of the Ricci tensor and the electromagnetic field. This result questions the mathematical foundations of both General Relativity (GR) and Gauge Theory (GT). In particular, the {\it Einstein} operator (6 terms) must be thus replaced by the adjoint of the {\it Ricci} operator (4 terms only) in the study of gravitational waves. \\

\noindent
{\bf KEY WORDS}\\

Elasticity, Electromagnetism, Thermoelectricity, Mach-Lippman analogy, Helmholtz analogy, Piezoelectricity, Photoelasticity, Riemann tensor, Ricci tensor, Einstein equations, Lie group, Lie pseudogroup, Janet sequence, Spencer sequence, Adjoint operator, Double duality.  \\

\newpage 
 
\noindent

{\bf 1) INTRODUCTION}  \\ 
 
  At the beginning of the last century G. Lippmann and H. von Helmholtz, who knew each other, were both looking for the possibility to interpret thermostatics and electric phenomena by exhibiting a common macroscopic mechanical origin through a kind of variational calculus similar to the one used in analytical mechanics for getting Euler-Lagrange equations. As a byproduct, it is not possible to separate the {\it Mach-Lippmann analogy} from the 
{\it Helmholtz analogy} that we now recall.\\

In analytical mechanics, if $L(t,q,\dot{q})$ is the {\it Lagrangian} of a 
mechanical system, one easily gets the {\it Hamiltonian} 
$H=\dot{q}\frac{\partial L}{\partial\dot{q}}-L$ where $t$ is time, $q$ 
represents a certain number of dependent variables or {\it
generalized position}, allowing to define the position of the various rigid 
bodies constituting the system (coordinates of center of gravity, relative 
angles, ...) and $\dot q$ is the derivative with respect to time or 
{\it generalized speed}. There are two ideas behind
such a construction. The first is to introduce the energy as in the
  movement of a point of mass $m$ with Cartesian coordinates
($x,y,z$ vertical) or ($x^1,x^2,x^3$ vertical) in the gravitational field 
$\vec{g}$ where 
$L=\frac{1}{2}m({\dot{x}}^2+{\dot{y}}^2+{\dot{z}}^2)-mgz$
  and thus 
$H=\frac{1}{2}m({\dot{x}}^2+{\dot{y}}^2+{\dot{z}}^2)+mgz$.
  The second is to take into account the well known Euler-Lagrange
  equations 
$\frac{d}{dt}\left(\frac{\partial L}{\partial\dot{q}}\right)-\frac{\partial L}{\partial
q}=0$ implied by the variational condition $\delta\displaystyle{\int}
  L(t,q,\dot{q})dt=0 $ and to obtain therefore~:
\begin{equation} \label{1} 
 \frac{dH}{dt}=\dot{q}\left(\frac{d}{dt}\left(\frac{\partial
  L}{\partial\dot{q}}\right)- \frac{\partial L}{\partial
  q}\right)-\frac{\partial L}{\partial t}=-\frac{\partial L}{\partial t} 
\end{equation} 
that is the conservation of energy along the trajectories whenever $L$ does
  not contain $t$ explicitly.\\

Similarly, in thermostatics, if $F$ is the {\it free energy} of a system
 at absolute temperature $T$, we may obtain, {\it in general}, the 
{\it internal energy} $U$ by the formula~: 
$U=F-T\frac{\partial F}{\partial T}$. We explain the underlying difficulty in the case of a perfect gas with 
pressure $P$, volume $V$ and entropy $S$ for one mole. The {\it first
principle} of thermostatics says that the sum of the exchange of work 
$\delta W=-PdV$ and the exchange of heat $\delta Q$ between the
system and its surrounding is a total differential $dU=\delta W+\delta Q$. 
Now, the {\it second principle} of thermostatics says that 
$\delta Q=TdS$ or equivalently that $\frac{\delta Q}{T}=dS$ is a total 
differential with absolute temperature as
integrating factor. Accordingly, we have 
$dU=-PdV+TdS$, a result giving $U$ as a function of $V$ and $S$. As $V$ 
{\it has a geometric meaning that} $S$ {\it does not possess}, engineers
use to do a Legendre transformation by introducing $F=U-TS$ in order to
 have  $dF=-PdV-SdT$ where $F$ is now a function of $V$ and $T$ that can be 
measured. It follows that $S=-\frac{\partial F}{\partial T}$ 
{\it in this situation} because $\delta W=-PdV$ does not contain $dT$. Of 
course, {\it contrary to} $S$, $T$ {\it can be measured} {\it though it does not seem to have a geometric meaning like} $V$. In
 general, the 1-form $\delta W$ depends linearly on the differentials of
 {\it all} the state
variables ($dV$ and $dT$ in our case) and {\it there is no reason at all to 
have again} $S=-\frac{\partial F}{\partial T}$. To avoid such a situation, 
{\it Helmholtz postulated the possibility for any system to choose
 " normal " state variables such that} $dT$ {\it
should not appear in} $\delta W$. Therefore, if 
one could introduce $V$ and $T$ on an equal geometric
footing, then $dF=-PdV-SdT$ should already contain, in a built-in manner,
 not only the first and second principle but also the well
defined possibility to recover $U$ from $F$ as before. In the case of 
continuum mechanics, $V$ must be replaced by the
deformation tensor, as we shall see later on, which is a function of the 
first order derivatives of the {\it actual} (Euler) position $x$ at
time $t$ with respect to the {\it initial} (Lagrange) position $x_0$ at
 time $t_0$. Accordingly, the idea of Helmholtz has been to
compare the relations $L\longrightarrow H$ and $F\longrightarrow U$ and to 
notice that they should become indeed similar if one could set 
$L=-F$ {\it and} $\dot{q}=T$ {\it for a certain} $q$. However, despite many 
attempts [7], nobody knows any variable $q$ such that its derivative with
respect to time should be the absolute temperature $T$ of the system under 
study. \\

 We now present the work done by Lippmann in a modern setting. The basic idea is to
compare two kinds of {\it conceptual experiments},
namely a Carnot cycle for a steam engine working
between the absolute temperatures $T_1$ and $T_2$ with $T_2>T_1$ on one
side, and a cycle of charge and discharge of a spherical condenser
(capacitor) of radius $r$, say a soap bubble, moving inbetween two plates at
constant electric potentials $V_1$ and $V_2$ with $V_2>V_1$ on the other
side ([18-21]).\\
In the first case, let the system receive the heat $Q_2>0$ from the hot 
source and the heat $Q_1<0$ from the cold source through corresponding 
isothermal evolutions, while receiving the work $W<0$ from the surroundings in a cycle completed by two adiabatic evolutions. \\
The vanishing of the cycle integral:  \\
\[    \oint(\delta W+\delta Q)=\oint dU=0   \]
coming from the first principle of thermostatics leads to the relation $W+Q_1+Q_2=0$. \\
Then, the vanishing of the cycle integral coming from the second principle of thermostatics:   \\
\[    \oint\frac{\delta Q}{T}=\oint dS=0   \]
leads to the {\it Clausius formula} and the computation of the {\it efficiency} $\nu$:   \\
\[   \frac{Q_1}{T_1}+\frac{Q_2}{T_2}=0  \,\,\, \Rightarrow \,\,\, \nu=\frac{ - W}{Q_2}=  \frac{Q_1+Q_2}{Q_2}=\frac{T_2-T_1}{T_2} > 0\]
Now, in the second case, {\it things are quite more subtle}. Recalling the formula $q=CV$ relating the charge $q$ to the potential $V$ of 
a condensor with $C=4\pi{\epsilon}_0r$ for a sphere of radius $r$, the electric energy should be:  \\
\[   E=\frac{1}{2}CV^2=\frac{1}{2}\frac{q^2}{C}=\frac{1}{2}qV \]  
{\it Whenever} $C$ {\it remains constant}, the exchange of work
done by the sources should be $\delta W'=Vdq$ because, 
{\it by definition, sources are at constant potential}, and we have $dE=qdV=Vdq=\delta W'$.\\

The situation is {\it completely different} in 
the second experiment because $C$ now depends on $r$ and we do not believe 
that Lippmann was very conscious about this 
fact. Let us suppose that the bubble receives the work $W'_2>0$ from the source
at potential $V_2$ for having its charge changing at constant potential
$V_2$ and similarly the work $W_1<0$ from the source at
constant potential $V_1$ for having its charge changing at constant
potential $V_1$, while receiving the (mechanical) work $W<0$
from the surroundings for changing $C$ in a cycle where the geometry of the 
system may vary (change of radius, distance, ...).
The problem is now to construct the cycle in order to be able to copy the 
procedure used for thermostatics. In the evolution at
constant potential we have $\delta W'=Vdq$, as already said, and therefore, 
comparing with $\delta Q=TdS$, {\it the remaining evolution
must be at constant charge}, a situation happily realized in the experiment 
proposed by Lippmann, during the transport of the bubble from one plate to the other. \\
Taking into account the expression $\delta W'=Vdq$ already introduced and allowing $C$ to vary (through $r$ in our case), 
we have now the formula: \\
\[ dE=-\frac{1}{2}V^2dC+Vdq=\delta W+\delta W'  \]
if we express $E$ as a function of $C$ and $q$. In our case 
$\delta W=-2\pi{\epsilon}_0V^2dr$ and the relation $q=CV$
plays the role of the relation $PV=RT$ existing for a perfect gas. \\
Copying the use of the first principle of thermostatics, the vanishing of the cycle integral provides:  \\
\[    \oint (\delta W+\delta W')=\oint dE=0   \,\,\,   \Rightarrow \,\,\,   W+W'_1+W'_2=0   \] 
Lippmann then notices that the conservation of entropy now 
becomes the conservation of charge and the vanishing of the cycle
integral provides:  \\
\[    \oint\frac{\delta W'}{V}=\oint dq=0 \,\,\,  \Rightarrow \,\,\, \frac{W'_1}{V_1}+\frac{W'_2}{V_2}=0  \,\,\, \Leftrightarrow \,\,\, \nu= \frac{ - W}{W'_2}= \frac{V_2 - V_1}{V_2} > 0  \]
analogous to the Clausius formula with similar {\it efficiency} $\nu$, a result called by Helmholtz ``{\it Principe de conservation de
  l'\'electricit\'e} '' or ``{\it Second principe de la th\'eorie des ph\'enom\`enes \'electriques} ". \\
One must notice the formula:  \\
 \[   dE=\frac{1}{2}V^2dC+qdV   \]
if we express $E$ as a function of $C$ and $V$. Also the analogue of the free energy should be $E-qV=-E$ 
expressed as a function of $C$ and $V$. Hence {\it it is not evident, at first sight, to know whether the more
  ``geometric" quantity is $q$ or $V$}. \\
  
Finally, the analogy between $T$ and $V$ in the corresponding 
``second principles" is clear and constitutes the Mach-Lippmann analogy. However, the reader may find strange that $T$, which is 
just defined up to a change of scale because of the existence of a reference absolute zero, should
be put in correspondence with $V$ which is defined up to an additive
constant. In fact, the formula for the spherical condenser (Gauss theorem)
is only true if {\it the potential at infinity is chosen to be zero}, as a 
zero charge on the sphere is perfectly detectable by counting the number of 
electrons on the surface. Accordingly, the two previous dimensionless 
ratios are perfectly well defined, independently of any
unit chosen for $T$ or $V$. However, such an analogy is perfectly coherent with the existence of 
thermocouples where the gradient of $T$ is proportional to the gradient of $V$,
that is we have for the electric field $\vec{E}=\eta(T)\vec{\nabla T}$ and 
the latter difficulty entirely disappears. \\

We recall that the {\it thermoelectric effect}, that is the existence of an 
electric current circulating in two different metal threads $A$ and $B$
with soldered ends at different temperatures $T_1$ and $T_2>T_1$, has 
been discovered in 1821 by the physicist Seebeck from the Netherlands. Also 
cutting one of the threads to set a condenser and integrating along the 
circuit, the difference of potential becomes:  \\
\[   V=\oint\vec{E}\cdot\vec{d\ell}=\int^{T_2}_{T_1}\left({\eta}_A(T)-{\eta}_B(T)\right)dT \]
Hence a thermocouple only works if $A\neq B$, $T_1\neq T_2$ and tables of 
coefficients can be found in the literature. It is the French
physicist Becquerel who got the idea in 1830 to use such a property for 
measuring temperature and Le Chatelier in 1905 who set up the platine 
thermocouple still used today. Meanwhile, J. Peltier proved that, when an 
electric current is passing in a thermocouple circuit with soldered joints
at the same temperature, then one of the joints absorbs heat while the
other produces heat. Also W. Thomson proved that an electric
current passing in a piece of homogeneous conductor in thermal equilibrium 
gives a difference of potential at the ends whenever they
are not at the same temperature. \\

The first criticism of the Mach-Lippmann analogy has been done by
E.W. Adler in 1907 [1]. His main claim is 
that, on the energy level (same Joule unit) the stored heat is like $cT$ 
while the stored electrical energy is $\displaystyle{\frac{1}{2}}CV^2$, 
a result showing a different behaviour. However, as stressed by Lippmann in 
his answer [19], a careful examination shows that things are different. 
Indeed, looking at corresponding concepts in this analogy, we have to set up on
the same level $\gamma=\frac{dS}{dT}$ and 
$C=\frac{dq}{dV}$ while $c=\frac{dQ}{dT}=\gamma T$. Hence, if we suppose
$C$ and similarly $\gamma$ to be constant (we already discussed this
assumption) we obtain respectively $\frac{1}{2}\gamma T^2$ and 
$\frac{1}{2}CV^2$ but it is more satisfactory (as we saw) to 
say that the analogy is on the formal level of the cyclic integrals. \\

We end this presentation of the Mach-Lippmann analogy with the main problem that it raises. From the special relativity of A. Einstein in 1905 [25] it 
is known that space cannot be separated from time and that one of the best examples is given by the relativistic formulation of EM.
Indeed, instead of writing down separately the first set of Maxwell equations for the electric field $\vec{E}$ and the magnetic field
$\vec{B}$ under their classical form, ne may introduce local coordinates $(x^1,x^2,x^3,x^4=ct)$ where $c$ is the speed of light and consider the 
2-form:  \\   
\[   F=B_1dx^2\wedge dx^3+B_2dx^3\wedge dx^1+B_3dx^1\wedge dx^2 +\frac{1}{c}E_1dx^1\wedge dx^4+\frac{1}{c}E_2dx^2\wedge dx^4+
\frac{1}{c}E_3dx^3\wedge dx^4   \]
in order to obtain:  \\
\[   \vec{\nabla}\cdot\vec{B}=0, \,\,\vec{\nabla}\wedge\vec{E}+\frac{\partial\vec{B}}{\partial t}=0 \,\,\Leftrightarrow  \,\,  dF=0  \]
where $d$ is the exterior derivative. \\
Similarly, introducing the electromagnetic potential $\vec{A}$ and the electric potential $V$ in the $1$-form 1-form $A=A_1dx^1+A_2dx^2+A_3dx^3+A_4dx^4$ where $A_4=-V/c$ is the time 
component, we obtain:   \\
\[  \vec{B}=\vec{\nabla}\wedge\vec{A}, \,\,\vec E=-\vec{\nabla}V - \frac{\partial\vec{A}}{\partial t} \,\,\, \Leftrightarrow \,\,\,  dA=F  \] 
though, surprisingly, $V$ has been introduced in thermostatics. Hence, even if we may accept and 
understand an analogy between $T$ and $V$, we cannot separate $V$ from 
$\vec{A}$ in the 4-potential $A$ and a good conceptual analogy
should be between $T$ and $A=(A_1,A_2,A_3,A_4)$. \\
The surprising fact is that almost nobody knows about the Mach-Lippmann analogy
but many persons are using it through finite element computations and thus any engineer working with finite elements knows 
that {\it elasticity}, {\it heat} and {\it electromagnetism}, though being quite different theories at first sight, are organized along the same scheme and cannot be separated because of the existence of the following three main couplings that we shall study with more details in the next Section.    \\

\noindent 
$\bullet$ {\it THERMOELASTICITY} (Elasticity/Heat):\\
When a bar of metal is heated, its length is increasing and, conversely,
its length is decreasing when it is cooled down. It is a perfectly 
{\it reversible phenomenon}. \\  

\noindent 
$\bullet$ {\it THERMOELECTRICITY} (Heat/Electromagnetism):\\ 
We have already spoken about this coupling which, nevertheless, can only be 
understood today within the framework of the phenomenological 
{\it Onsager relations} for {\it irreversible phenomena}. \\
Hence we discover that the Mach-Lippmann analogy must be set up in a clear 
picture of the analogy existing between elasticity, heat an
electromagnetism that must also be coherent with the above couplings. \\

\noindent 
$\bullet$ {\it PIEZOELECTRICITY, PHOTOELASTICITY} (Elasticity/Electromagnetism):\\
When a crystal is pinched between the two plates of a condenser, it produces 
a difference of potential between the plates and conversely, in a purely 
reversible way. Piezoelectric lighters are of common use in industry. \\
Similarly, when a transparent homogeneous isotropic 
dielectric is deformed, piezoelectricity cannot appear 
but the index of refraction becomes different along the three orthogonal proper
directions common to both the strain and stress tensors. Here we recall
that a material is called ``{\it homogeneous}" if a property does not
depend on the point in the material and it is called ``{\it isotropic}" if
a property does not depend on the direction in the material. Accordingly, a 
light ray propagating along one of these directions may have its electric 
field decomposed along the two others and the two components propagate with 
different speeds. Hence, after crossing the material, they recompose with 
production of an interference pattern, a fact leading to optical 
birefringence. Such a property has been used in order to get information on the
stress inside the material, say a bridge or a building, by using reduced 
transparent plastic models. This phenomenon was discovered by
Brewster in 1815 but the phenomenological law that we shall prove in the next section, was proposed independently by 
F.E. Neumann and J.C. Maxwell in 1830. Until recently one used to rely on
the mathematical formulation proposed by P\"{o}ckels in 1889 but modern
versions can easily be found today in the literature. \\

\newpage

\noindent
{\bf 3) ELASTICITY VERSUS ELECTROMAGNETISM}   \\

The rough idea is to make the constitutive law of an homogeneous isotropic dielectric $\vec{D}=\epsilon
\vec{E}$ where $\vec{D}$ is the electric induction and $\epsilon=\epsilon_0(1+\chi)$, $\epsilon_0$
being the vacuum value (universal constant) of the dielectric constant, such that {\it the dielectric susceptibility $\chi$ now depends on the 
deformation (or stress) tensor in each direction}. Keeping the constitutive relation $\vec{H}=\frac{1}{\mu}\vec{B}$ where $\vec{H}$ is
the magnetic induction and $\mu=\mu_0$ the vacuum value (universal constant) of the magnetic constant, as we have no magnetic polarization in 
the medium, it is well known that $\epsilon_0\mu_0c^2=1$ and thus $\epsilon\mu c^2=n^2$ where $n$ is the
index of refraction such that $n^2=(1+\chi)$, a result leading to the Maxwell-Neumann formula $ \sigma_1-\sigma_2=k\lambda / eC $
that we shall demonstrate and apply to the study of a specific beam. In this formula $\sigma_1,\sigma_2$ are the two eigenvalues of the symmetric stress
tensor along directions orthogonal to the ray, $k$ is a relative integer fixing the lines of interference, $\lambda$ is the wave length, $e$ is
the thickness of the transparent beam and $C$ is the photoelastic constant of the material. \\

\noindent  

With more details, the infinitesimal deformation tensor of elasticity theory is equal to half of the Lie derivative $\Omega=({\Omega}_{ij}={\Omega}_{ji})={\cal{L}}(\xi)\omega$ of the euclidean metric $\omega$ with respect to the displacement vector $\xi$. Hence, a general quadratic lagrangian may contain, apart from its standard purely elastic or electrical parts well known by engineers in finite element computations, a coupling part  $c^{ijk}{\Omega}_{ij}E_k$ where $E=(E_k)$ is the electric field. The corresponding induction $D=(D^k)$ becomes: \\
\[ D^k_0=\epsilon E^k  \longrightarrow   D^k=D^k_0+c^{ijk}{\Omega}_{ij}  \] 
and is therefore modified by an electric  polarization $P^k=c^{ijk}{\Omega}_{ij}$, brought by the deformation of the medium. In all these formulas and in the forthcoming ones the indices are raised or lowered by means of the euclidean metric. If this medium is homogeneous, the components of the $3$-tensor $c$ are constants and the corresponding coupling, called {\it piezoelectricity}, is only existing if the medium is non-isoptropic (like a crystal), because an isotropic $3$-tensor vanishes identically.\\
In the case of an homogeneous isotropic medium (like a transparent plastic), one must push the coupling part to become cubic by adding $\frac{1}{2}d^{ijkl}{\Omega}_{ij}E_kE_l$ with $d^{ijkl}=\alpha {\omega}^{ij}{\omega}^{kl}+\beta {\omega}^{ik}{\omega}^{jl}+\gamma {\omega}^{il}{\omega}^{jk}$ from Curie's law. The corresponding coupling, called {\it photoelasticity}, has been discovered by T.J. Seebeck in 1813 and D. Brewster in 1815. With $\delta=\beta + \gamma$, the new electric induction is:\\
\[   D^k_0=\epsilon E^k \longrightarrow D^k=D^k_0+(\alpha \hspace{1mm}tr(\Omega){\omega}^{kr}+\delta{\omega}^{ik}{\omega}^{jr}{\Omega}_{ij})E_r  \]
As $\Omega$ is a symmetric tensor, we may choose an orthogonal frame at each point of the medium in such a way that the deformation tensor becomes diagonal with $\Omega=({\Omega}_1,{\Omega}_2,{\Omega}_3)$ where the third direction is orthogonal to the elastic plate. We get: \\
\[  D^i=D^i_0+(\alpha \hspace{1mm}tr(\Omega)+\delta {\Omega}_i)E^i  \]
for $i=1,2$ without implicit summation and there is a change of the dielectric constant $\epsilon \longrightarrow \epsilon +\alpha \hspace{1mm}tr(\Omega)+\delta {\Omega}_i$ along each proper direction in the medium, corresponding to a change $n \longrightarrow n_i$ of the refraction index. As there is no magnetic property of the medium and $\Omega \ll 1$, we obtain in first approximation:\\
\[  \epsilon {\mu}_0c^2=n^2  \Longrightarrow  n_1^2-n_2^2\simeq 2n(n_1-n_2)={\mu}_0c^2\delta({\Omega}_1-{\Omega}_2) \Longrightarrow n_1-n_2 \sim {\Omega}_1-{\Omega}_2  \]
where ${\mu}_0$ is the magnetic constant of vacuum, $c$ is the speed of light in vacuum and $n$ is the refraction index. The speed of light in the medium becomes $c/n_i$ and therefore depends on the polarization of the beam. As the light is crossing the plate of thickness $e$ put between two polarized filters at right angle, the entering monochromatic beam of light may be decomposed along the two proper directions into two separate beams recovering together after crossing with a time delay equal to:
\[     e/(c/n_1) -e/(c/n_2)=(e/c)(n_1-n_2)  \]
providing interferences and we find back the Maxwell phenomenological law of 1850: \\
\[    {\Omega}_1-{\Omega}_2\sim  {\sigma}_1-{\sigma}_2 =\frac{k\lambda}{eC}  \]
where $\sigma$ is the stress tensor, $k$ is an integer, $\lambda$ is the wave length of the light used and $C$ is the photoelastic constant of the medium ivolved in the experience.\\

Looking at the picture, let $F$ be the vertical downwards force acting on the upper left side of the beam like on the picture, at a distance $D$ from the center of the vertical beam on the right. We may consider this vertical beam as a dense sheaf of juxtaposed thin beams with young modulus $E$. Choosing orthogonal axes $(Oxyz)$ such that $Ox$ is horizontal towards the right with origin $O$ in the geometric center of the vertical beam on the right which has a Thickness $e=2a$ and a  width of $2b$ with the vertical axis $Oy$ passing in the center of the beam. If $F$ should be apllied along $Oy$, according to Hooke's law there should be a vertical compression of the beam providing a deformation roughly equal to ${\epsilon}'= - F/(4abE)$ and a (negative because compression) stress ${\sigma}'= E{\epsilon}'= - F/4ab$. However, $F$ is applyed at a distance $D$ of the axis $Oy$ and gives a couple $C=FD$ which should be, by itself, bringing the half right part of the beam ( $x\geq 0$) in extension while the half left part ($x\leq 0$) is in compression. Using a classical assumption usually done on beams we may suppose that the horizontal plane sections orthogonal to the central axis $Oy$ of the beam stay plane surfaces turning counterclockwise by a small angle $\theta$ that we shall determine by integration on all the small thin beams of the bunch. The stress ${\sigma}"$ acting on the surface $dS=edx=2adx$ is producing a small force $dF={\sigma}"dS=2a{\sigma}" dx$. However, a fiber at distance x from the axis has a length increased by $\theta x$ and there is a resulting deformation ${\epsilon}"=Kx$ such that ${\sigma}"=E{\epsilon}"$ on each thin constitutive fiber. The resulting (direct sense) couple produced is equal to $dM=x (2aKExdx)$ in such a way that we have the equilibrium equation for couples:  \\
\[ M=FD={\int}^{+b}_{-b}(2aKEx^2)dx=\frac{4}{3}ab^3KE   \hspace{5mm}       \Rightarrow   \hspace{5mm}  K=(3FD)/(4ab^3E)  \]
We obtain therefore ${\sigma}"=EKx=(3FDx)/(4ab^3)$ with ${\sigma}"\geq 0$ whenever $x\geq 0$ (extension). Using the correct negative sign for the 
stress ${\sigma}"$, we finally obtain $\sigma = {\sigma}^{yy}= {\sigma}' + {\sigma}"= (3FDx)/(4ab^3) - F/(4ab)$ in such a way that $\sigma \leq 0$ when $D=0$ and $\sigma = 0$ when $x=b^2/3D> 0$, a result not evident at first sight. In addition, it is clear by symmetry that $x,y,z$ are proper directions and that ${\sigma}^{xx}={\sigma}^{zz}=0$ because no force is acting on the faces of the beam. We obtain therefore the very simple Maxwell law $\sigma=k\lambda/eC$. Accordingly, the (almost !) central black line corresponds to $\sigma =0$ and has abcissa $x=b^2/3D >0$. Finally, the distance $d$ between two lines is such that $k$ is modified by $1$, that is $d=(2\lambda b^3)/(3FDC)$, allows to determine the photoelastic constant of the material. \\
The study of the upper horizontal part of the beam is more delicate. With axis $Oy$ in the middle section, starting under the force $F$ and axix $Ox$ upward, we have ${\sigma}^{yy}=(3Fxy)/(2eb^3)$ to compensate the couple $Fy$ but {\it now} we have a shear stress ${\sigma}^{xy}=F/(2eb)$ upward to compensate $F$ which is downward. The characteristic polynomial is $det(\sigma -\lambda \omega)={\lambda}^2-{\sigma}^{yy}\lambda -({\sigma}^{xy})^2=(\lambda - {\sigma}^1)(\lambda - {\sigma}^2)=0$ and thus $({\sigma}^1-{\sigma}^2)^2= ({\sigma}^1+{\sigma}^2)^2- 4{\sigma}^1{\sigma}^2=({\sigma}^{yy})^2 + ({\sigma}^{xy})^2 > 0$ cannot vanish. Therefore the line "$k=0$" cannot exist. As for the lines "$k=\pm 1$", we must have after substitution $((3Fxy)/2eb^3))^2 + (F/2eb)^2=(\lambda/eC)^2$ and we need to have thus $F< (2\lambda b)/C$ or equivalently $d>b^2/3D$, a result simply leading to the hyperbola $xy=cst$, a property that can be checked on the picture but cannot be imagined.  \\

We have thus explained, in a perfectly coherent way with the picture, why the interference lines are parallel and equidistant from each other in the right vertical part of the beam, on both sides of an ({\it almost}) central line which, {\it surprisingly}, stops at the upper and lower corner, even though, by continuity, we could imagine that it could be followed in the upper and lower horizontal parts of the beam. Also, we understand now the reason for which the lines in these parts of the beam look like symmetric hyperbolas.  \\

 \newpage

\begin{figure}[p]

\includegraphics[scale=0.8]{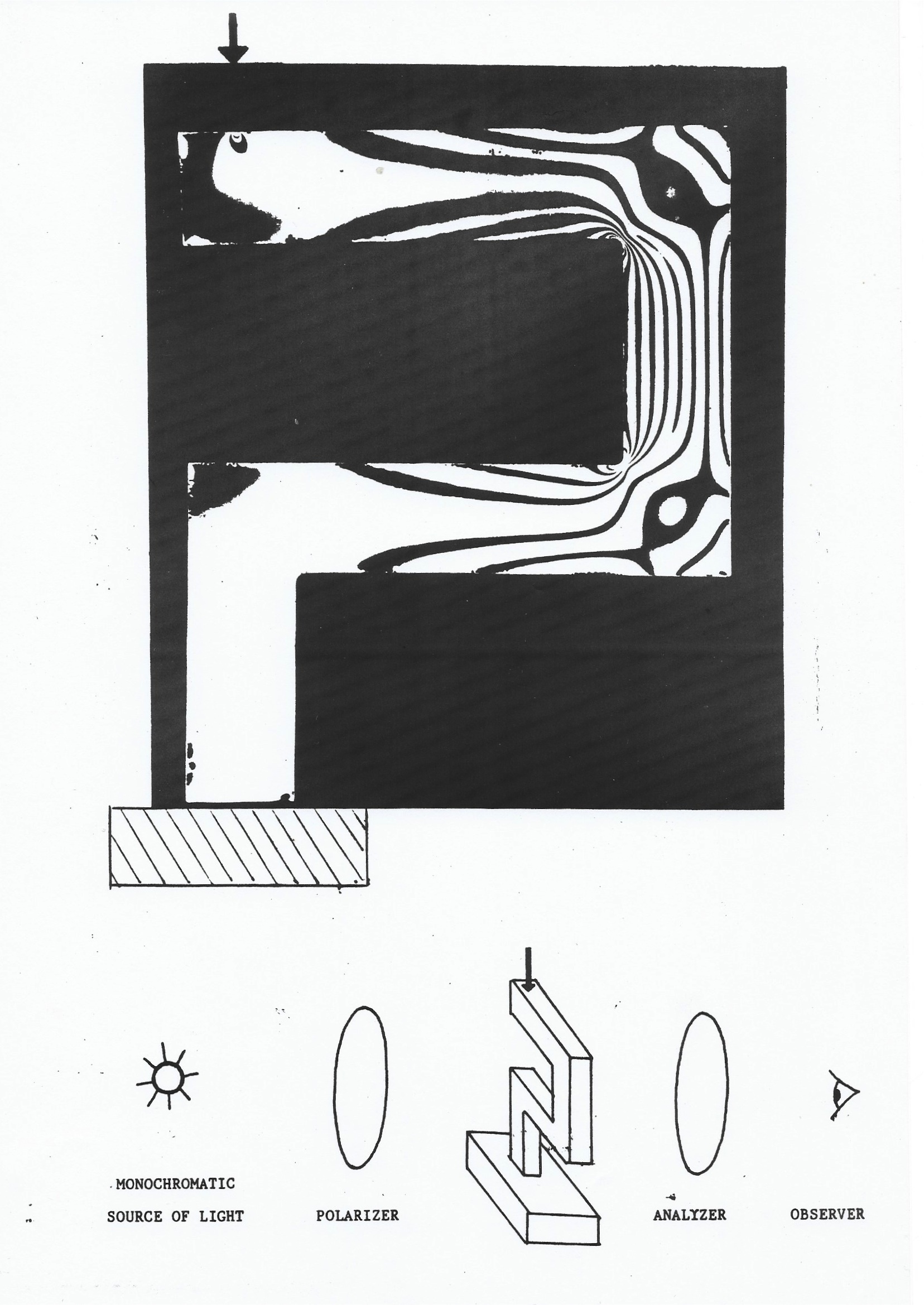}

 \end{figure}

 \newpage

This result proves, {\it without any doubt for anybody doing this experiment}, that the deformation $\Omega={\cal{L}}(\xi)\omega$ and the electromagnetic field $F=dA$, using standard notations in the space-time formulation of electromagnetism, {\it must be on equal footing} in a lagrangian formalism. However, as $\Omega\in S_2T^*$ is in the {\it Janet sequence} based on the work of E. Vessiot in $1903$ (Compare [50] to [27-30]) and $F\in {\wedge}^2T^*$ {\it cannot} appear at this level as we shall see, the main purpose of this paper is to prove that {\it another differential sequence must be used}, namely the {\it Spencer sequence}. The idea has been found {\it totally independently}, by the brothers E. and F. Cosserat in $1909$ ([11]) for revisting elasticity theory and by H. Weyl in $1916$ ([51]) for revisiting electromagnetism by using the conformal group of space time, but the first ones were only dealing with the {\it translations} and {\it rotations} while the second was only dealing with the {\it dilatation} and the non-linear {\it elations} of this group, with no real progress during the last hundred years.  \\

Extending the space $(x^1,x^2,x^3)$ or $(x,y,z)$ to space-time 
$(x^1,x^2,x^3,x^4=ct)$ as before, the speed is now extended from
$(v^1,v^2,v^3)$ to $(v^1,v^2,v^3,c)$ along the derivative with respect to 
time, with $v/c \ll 1$, while the motion
$x=x_0+\xi(x_0,t)$ is extended to $t=t_0+ cst$ in order to compare
``slices" of space at the same ``time". Accordingly the
deformation tensor $\epsilon$, which is dimensionless, is extended by
$\left(\epsilon_{i4}=\epsilon_{4i}\right)=\frac{1}{2}\left(\frac{v^1}{c},
\frac{v^2}{c},\frac{v^3}{c},0\right)$ while the symmetric
stress tensor $\sigma^{ij}=\sigma^{ji}$ becomes $\sigma^{ij}-\rho v^iv^j$ 
(Euler theorem) and is extended by setting ${\sigma}_{i4}= - {\sigma}^{i4}=\rho v^ic$, $\sigma^{44}=\sigma_{44}=-\rho c^2$ 
where $\rho$ is the mass per unit volume. Dealing with the rest-frame and 
using the (small) dilatation relation $\rho={\rho}_0(1-tr\epsilon)$ in which
$\rho_0$ is the value of $\rho$ in the initial position where the body is 
supposed to be homogeneous, isotropic and unstressed, that
is, $\rho_0$ is supposed to be a constant. The Hooke law is now extended by setting:  \\
\[   \sigma_{i4}=2{\rho}_0c^2{\epsilon}_{i4},\,\,{\sigma}_{44}+{\rho}_0c^2= {\rho}_0c^2 tr \epsilon \] 
in a way compatible with the conservation of mass and we suddenly discover 
that {\it there is no conceptual difference between the Lam\'e constants} $(\alpha, \beta)$ (do not confuse the notations) of elasticity 
and {\it the magnetic constant} $\mu$ {\it on one side} 
({\it space}) or {\it the mass per unit volume $\rho$ and the dielectric 
constant} $\epsilon$ (time) {\it on the other side}, all these coupling constants being measured in the
reference state in which the body (like vacuum) is homogeneous
and isotropic (the index ``zero" is omitted for simplicity). This result is perfectly coherent ``{\it a posteriori} " with the analogy existing between
the well known formulas for {\it the speed} $v_T$ {\it of transverse elastic waves, the speed $v_L$ of longitudinal elastic waves} or
{\it the speed $v$ of light waves} propagating in a homogeneous isotropic medium, as we have indeed:  \\
\[   v_T=\sqrt{\frac{\beta}{\rho}}, \hspace{1cm} v_L=\sqrt{\frac{\alpha+2\beta}{\rho}},
\hspace{1cm} v=\sqrt{\frac{1}{\epsilon\mu}}=\sqrt{\frac{1/\mu}{\epsilon}}=\frac{c}{n}\]
We now understand that couplings are in fact more general constitutive laws taking into account the tensorial nature of the 
various terms involved through the {\it Curie principle}. \\

\newpage

\noindent
{\bf 2) GENERAL RELATIVITY VERSUS GAUGE THEORY}  \\

Let $A$ be a {\it unitary ring}, that is $1,a,b\in A \Rightarrow a+b,ab \in A, 1a=a1=a$ and even an {\it integral domain} ($ab=0\Rightarrow a=0$ or $b=0$) with {\it field of fractions} $K=Q(A)$. However, we shall not always assume that $A$ is commutative, that is $ab$ may be different from $ba$ in general for $a,b\in A$. We say that $M={}_AM$ is a {\it left module} over $A$ if $x,y\in M\Rightarrow ax,x+y\in M, \forall a\in A$ or a {\it right module} $M_B$ over $B$ if the operation of $B$ on $M$ is $(x,b)\rightarrow xb, \forall b\in B$. If $M$ is a left module over $A$ and a right module over $B$ with $(ax)b=a(xb), \forall a\in A,\forall  b\in B, \forall x\in M$, then we shall say that $M={ }_AM_B$ is a {\it bimodule}. Of course, $A={ }_AA_A$ is a bimodule over itself. We define the {\it torsion submodule} $t(M)=\{x\in M\mid \exists 0\neq a\in A, ax=0\}\subseteq M$ and $M$ is a {\it torsion module} if $t(M)=M$ or a {\it torsion-free module} if $t(M)=0$. We denote by $hom_A(M,N)$ the set of morphisms $f:M\rightarrow N$ such that $f(ax)=af(x)$ and set $M^*=hom_A(M,A)$. We finally recall that a sequence of modules and maps is exact if the kernel of any map is equal to the image of the map preceding it ([6, 13, 23, 28, 32, 33, 46] are good references for commutative and homological algebra). \\

When $A$ is commutative, $hom(M,N)$ is again an $A$-module for the law $(bf)(x)=f(bx)$ as we have $(bf)(ax)=f(bax)=f(abx)=af(bx)=a(bf)(x)$. In the non-commutative case, things are more complicate and, given ${}_AM$ and ${}_AN_B$, then $hom_A(M,N)$ becomes a right module over $B$ for the law $(fb)(x)=f(x)b$. \\

\noindent
{\bf DEFINITION 2.1}: A module $F$ is said to be {\it free} if it is isomorphic to a (finite) power of $A$ called the {\it rank} of $F$ over $A$ and denoted by $rk_A(F)$ while the rank $rk_A(M)$ of a module $M$ is the rank of a maximum free submodule $F\subset M$. It follows from this definition that $M/F$ is a torsion module. In the sequel we shall only consider {\it finitely presented} modules, namely {\it finitely generated} modules defined by exact sequences of the type $F_1 \stackrel{d_1}{\longrightarrow} F_0 \stackrel{p}{\longrightarrow} M\longrightarrow 0$ where $F_0$ and $F_1$ are free modules of finite ranks $m_0$ and $m_1$ often denoted by $m$ and $p$ in examples. A module $P$ is called {\it projective} if there exists a free module $F$ and another (projective) module $Q$ such that $P\oplus Q\simeq F$.\\

\noindent
{\bf PROPOSITION 2.2}: For any short exact sequence $0\rightarrow M' \stackrel{f}{\longrightarrow} M \stackrel{g}{\longrightarrow} M" \rightarrow 0$, we have the important relation $rk_A(M)=rk_A(M')+rk_A(M")$, even in the non-commutative case. As a byproduct, if $M$ admits a finite length free {\it resolution}  \,$ ... \, \stackrel{d_2}{\longrightarrow} F_1 \stackrel{d_1}{\longrightarrow} F_0 \stackrel{p}{\longrightarrow} M \rightarrow 0$, we may define the {\it Euler-Poincar\'{e} characteristic} \,${\chi}_A(M)={\sum}_r(-1)^r rk_A(F_r)=rk_A(M)$.  \\

We now turn to the operator framework with modules over the ring $D=K[d_1, ...,d_n]=K[d]$ of differential operators with coefficients in a differential 
field $K$ with $n$ commuting derivations $({\partial}_1,...,{\partial}_n)$, also called $D$-modules. Then $D$ is a differential bimodule over itself ([5, 10, 16, 32, 33, 45,47] are good references for differential modules and algebraic analysis) (More generally, [27-30, 32, 33, 48] are good references for the formal theory of systems of partial differential equations). \\

\noindent
{\bf DEFINITION 2.3}: If a differential operator $\xi \stackrel{\cal{D}}{\longrightarrow} \eta$ is given, a {\it direct problem} is to find generating {\it compatibility conditions} (CC) as an operator $\eta \stackrel{{\cal{D}}_1}{\longrightarrow} \zeta $ such that ${\cal{D}}\xi=\eta \Rightarrow {\cal{D}}_1\eta=0$. Conversely, given $\eta \stackrel{{\cal{D}}_1}{\longrightarrow} \zeta$, the {\it inverse problem} will be to look for $\xi \stackrel{\cal{D}}{\longrightarrow} \eta$ such that ${\cal{D}}_1$ generates the CC of ${\cal{D}}$ and we shall say that ${\cal{D}}_1$ {\it is parametrized by} ${\cal{D}}$ {\it if such an operator} ${\cal{D}}$ {\it is existing}. \\

Introducing the morphism $\epsilon: M \rightarrow M^{**}$ such that $\epsilon (m)(f)=f(m), \forall m\in M, \forall f\in M^*$ and defining the differential module $N$  from $ad({\cal{D}}_1)$ exactly like we defined the differential module $M$ from ${\cal{D}}$, we finally notice that any operator is the adjoint of a certain operator because $ad(ad(P))=P, \forall P \in D$ and we get ([9, 32, 33, 38, 42, 43]):  \\

\noindent
{\bf THEOREM 2.4}: ({\it reflexivity test}) In order to check whether $M$ is reflexive or not, that is to find out a parametrization if $t(M)=0$ which {\it can be again parametrized}, the test has 5 steps which are drawn in the following diagram where $ad({\cal{D}})$ generates the CC of $ad({\cal{D}}_1)$ and ${\cal{D}}_1'$ generates the CC of ${\cal{D}}=ad(ad({\cal{D}}))$ while $ad({\cal{D}}_{-1})$ generates the CC of $ad({\cal{D}})$ and ${\cal{D}}'$ generates the CC of ${\cal{D}}_{-1}$:  \\
\[  \begin{array}{rcccccccl}
 & & & & & {\eta}'     & &  {\zeta}' &\hspace{15mm} 5  \\
 & & & &  \stackrel{{\cal{D}}'}{\nearrow}   & & \stackrel{{\cal{D}}'_1}{\nearrow} &  &  \\
4 \hspace{15mm}&\phi & \stackrel{{\cal{D}}_{-1}}{\longrightarrow}& \xi  & \stackrel{{\cal{D}}}{\longrightarrow} &  \eta & \stackrel{{\cal{D}}_1}{\longrightarrow} & \zeta &\hspace{15mm}   1  \\
 &  &  &  &  &  &  &  &  \\
 &  &  &  &  &  &  &  &  \\
 3 \hspace{15mm}& \theta &\stackrel{ad({\cal{D}}_{-1})}{\longleftarrow}& \nu & \stackrel{ad({\cal{D}})}{\longleftarrow} & \mu & \stackrel{ad({\cal{D}}_1)}{\longleftarrow} & \lambda &\hspace{15mm} 2
  \end{array}  \]
\[{\cal{D}}_1 \,\,\,parametrized \,\,\,by \,\,\,{\cal{D}} \Leftrightarrow {\cal{D}}_1={\cal{D}}'_1  \Leftrightarrow ext^1(N)=0  \Leftrightarrow  \epsilon \,\,\, injective  \Leftrightarrow t(M)=0\] 
\[{\cal{D}} \,\,\,parametrized \,\,\,by \,\,\,{\cal{D}}_{-1} \Leftrightarrow {\cal{D}}={\cal{D}}' \Leftrightarrow ext^2(N)=0 \Leftrightarrow \epsilon \,\,\, surjective  \hspace{17mm}  \]

\noindent
{\bf COROLLARY 2.5}: In the differential module framework, if $F_1 \stackrel{{\cal{D}}_1}{\longrightarrow} F_0 \stackrel{p}{\longrightarrow} M \rightarrow 0$ is a finite free presentation of $M=coker({\cal{D}}_1)$ with $t(M)=0$, then we may obtain an exact sequence $F_1 \stackrel{{\cal{D}}_1}{\longrightarrow} F_0 \stackrel{{\cal{D}}}{\longrightarrow} E $ of free differential modules where ${\cal{D}}$ is the parametrizing operator. However, there may exist other parametrizations $F_1 \stackrel{{\cal{D}}_1}{\longrightarrow} F_0 \stackrel{{\cal{D}}'}{\longrightarrow} E' $ called {\it minimal parametrizations} such that $coker({\cal{D}}')$ is a torsion module and we have thus $rk_D(M)=rk_D(E')$.  \\
 
These results have been used in {\it control theory} and it is now known that a control system is {\it controllable} if and only if it is parametrizable ( (See [24,32,33,52] for more details). Keeping the same "{\it operational} " notations for simplicity, we may state ([32], p 638-650):  \\

\noindent
{\bf DEFINITION 2.6}: We say that ${\cal{D}}:\xi \rightarrow \eta$ admits a ({\it generalized}) {\it lift} ${\cal{P}}:\eta \rightarrow \xi$ if ${\cal{D}}\circ {\cal{P}}\circ {\cal{D}}={\cal{D}}$. The differential module determined by ${\cal{D}}$ is projective if and only if ${\cal{D}}$ admits a lift. \\

The following results have never been used for aplications:  \\

\noindent
{\bf LEMMA 2.7}: If ${\cal{D}}$ admits a lift, then $ad({\cal{D}})$ also admits a lift.  \\

\noindent
{\bf PROPOSITION 2.8}: If ${\cal{D}}$ parametrizes ${\cal{D}}_1$ and admits a lift ${\cal{P}}$, then ${\cal{D}}_1$ admits a lift ${\cal{P}}_1$ and we have the striking {\it Bezout identity} ${\cal{D}}\circ {\cal{P}}+{\cal{P}}_1 \circ {\cal{D}}_1= id_{\eta}$. Accordingly, the corresponding differential sequence, which is formally exact by definition, is also locally exact. \\

\noindent
{\bf COROLLARY 2.9}: If ${\cal{D}}_1$ generates the CC of ${\cal{D}}$ and both operators admit lifts, then $ad({\cal{D}})$ generates the CC of $ad({\cal{D}}_1)$.  \\

\noindent
{\bf EXAMPLE 2.10}: With $n=2, m=2, q=1, a\in K=\mathbb{Q}(x^1, x^2), D=K[d_1,d_2], D\eta=D{\eta}^1+D{\eta}^2$ and $\Phi\equiv d_1{\eta}^1+d_2{\eta}^2-a {\eta}^1$ we shall prove that $M_1=D\eta/D\Phi$ is torsion-free but not projective when ${\partial}_2a=0$ and projective but not free when ${\partial}_2 a\neq 0$, for example when $a=x^2$. Multiplying $\Phi$ by a test function $\lambda$ and integrating by parts formally the equation ${\cal{D}}_1\eta=\zeta$, we get the operator $ad({\cal{D}}_1)$ in the form:  \\
\[  - d_1 \lambda - a \lambda={\mu}^1,    \,\,\,\,   - d_2 \lambda={\mu}^2 \,\,\, \Rightarrow \,\,\, ({\partial}_2a)\lambda = d_1{\mu}^2-d_2{\mu}^1+a {\mu}^2   \]
\noindent
$\bullet$ ${\partial}_2a=0$: We get the only generating CC $ d_1{\mu}^2-d_2{\mu}^1+a {\mu}^2 =0$ and $ad({\cal{D}}_1)$ is not injective. There is therefore no lift and thus no splitting. Multilying by a test function $\phi$ and integrating by parts, we obtain the parametrization ${\cal{D}}:\phi \rightarrow \xi$ in the form $d_2 \phi=y^1, - d_1\phi +a \phi=y^2$ which is not injective. The corresponding sequence $D\stackrel{{\cal{D}}_1}{\longrightarrow} D^2 \stackrel{{\cal{D}}}{\longrightarrow} D $ with differential modules and its formal adjoint are both formally exact.  \\  
\noindent
$\bullet$ ${\partial}_2a\neq 0$: The situation is now {\it totally different}. In order to prove this, if we suppose that $a=x^2 $, we get the lift 
$\lambda=d_1{\mu}^2-d_2{\mu}^1+x^2{\mu}^2$ with adjoint $d_2\zeta={\eta}^1, -d_1\zeta+x^2\zeta={\eta}^2$ providing a lift ${\cal{P}}_1$ for ${\cal{D}}_1$. Substituting, we obtain two second order CC ${\nu}^1$ and ${\nu}^2$ satisfying the only CC $d_1{\nu}^2-d_2{\nu}^1+x^2{\nu}^2=0$. Multiplying these two CC by the test functions ${\xi}^1$ and ${\xi}^2$ and integrating by parts, we finally obtain the involutive parametrizing operator 
${\cal{D}}$ in the form:  \\
\[  d_{12}{\xi}^1+d_{22}{\xi}^2-x^2d_2{\xi}^1-2{\xi}^1={\eta}^1, \,\,\, - d_{11}{\xi}^1- d_{12}{\xi}^2+2x^2d_1{\xi}^1+x^2d_2{\xi}^2-(x^2)^2{\xi}^1- {\xi}^2={\eta}^2  \]
We get the long formally and locally exact differential sequence $  0 \rightarrow D \stackrel{{\cal{D}}_{-1}}{\rightarrow} D^2 \stackrel{{\cal{D}}}{\rightarrow}D^2 \stackrel{{\cal{D}}_1}{\rightarrow} D \rightarrow 0 $ and invite the reader to find a lift for the 
central operator as an exercise.\\

\noindent
{\bf EXAMPLE 2.11}: When $n=3$, the $div $ operator can be parametrized by the $curl$ operator which can be itself parametrized by the $grad$ operator. However, using $({\xi}^1,{\xi}^2, {\xi}^3=0)$, we may obtain the new minimal parametrization $ - {\partial}_3{\xi}^2={\eta}^1, {\partial}_3{\xi}^1={\eta}^2, {\partial}_1{\xi}^2-{\partial}_2{\xi}^1={\eta}^3\Rightarrow {\partial}_1{\eta}^1+{\partial}_2{\eta}^2+{\partial}_3{\eta}^3=0$ which cannot be again parametrized ([41, 43]).  \\

\noindent 
{\bf EXAMPLE 2.12}: {\it Parametrization of the Cauchy stress equations}.  \\ 
We shall consider the cases $n=2,3,4$ but the case $n$ arbitrary could be treated as well.  \\

\noindent
$\bullet$ $n=2$: The stress equations become ${\partial}_1{\sigma}^{11}+{\partial}_2{\sigma}^{12}=0, {\partial}_1{\sigma}^{21}+{\partial}_2{\sigma}^{22}=0$. Their second order parametrization ${\sigma}^{11}={\partial}_{22}\phi, {\sigma}^{12}={\sigma}^{21}=-{\partial}_{12}\phi, {\sigma}^{22}={\partial}_{11}\phi$ has been provided by George Biddell Airy in 1863 ([2]) and we shall thus denote by $Airy: \phi \rightarrow \sigma$ the corresponding operator. We get the linear second order system with formal notations:  \\
\[ \left\{  \begin{array}{rll}
{\sigma}^{11} & \equiv d_{22}\phi =0 \\
-{\sigma}^{12} & \equiv d_{12}\phi =0 \\
{\sigma}^{22} & \equiv d_{11}\phi=0
\end{array}
\right. \fbox{ $ \begin{array}{ll}
1 & 2   \\
1 & \bullet \\  
1 & \bullet  
\end{array} $ } \]
which is involutive with one equation of class $2$, $2$ equations of class $1$ and it is easy to check that the $2$ corresponding first order CC are just the stress equations. Now, multiplying the Cauchy stress equations respectively by test functions ${\xi}^1$ and ${\xi}^2$, then integgrating by parts, we discover that (up to sign and a factor $2$) the {\it Cauchy operator} is the formal adjoint of the {\it Killing operator} defined by ${\cal{D}}\xi={\cal{L}}(\xi)\omega =\Omega \in S_2T^*$, introducing the standard Lie derivative of the (non-degenerate) euclidean metric $\omega$ with respect to $\xi$ and using the fact that we have ${\sigma}^{ij}{\Omega}_{ij}={\sigma}^{11}{\Omega}_{11} +2 {\sigma}^{12}{\Omega}_{12}+ {\sigma}^{22}{\Omega}_{22}$ because we have supposed that ${\sigma}^{12}={\sigma}^{21}$ and we shall say, with a slight abuse of language, that $Cauchy = ad (Killing)$. In order to apply the above parametrization test, we have to look for the CC ${\cal{D}}_1$ of ${\cal{D}}$. In arbitrary dimension $n$, introducing the Riemann tensor ${\rho}^k_{l,ij}$ with $n^2(n^2-1)/12$ components of a general metric $\omega$ with $det(\omega)\neq 0$ and linearizing it over a given non-degenerate constant metric or, more generally, over a metric with constant Riemaniann curvature, we obtain the second order {\it Riemann operator} ${\Omega}_{ij} \rightarrow R^k_{l,ij}$. When $n=2$ and $\omega$ is the euclidean metric, we get a single component that can be choosen to be the scalar 
curvature $R= d_{11}{\Omega}_{22}+d_{22}{\Omega}_{11} - 2 d_{12}{\Omega}_{12}$. Multiplying by a test function $\phi$ and integrating by parts, we obtain $Airy = ad (Riemann)$ and notice that:  \\
{\it There is no relation at all between the Airy stress function} $\phi$ {\it and the deformation} $\Omega$ {\it of the metric} $ \omega$.  \\

\noindent
$\bullet \hspace{3mm} n=3$: Things become quite more delicate when we try to parametrize the $3$ PD equations: \\
\[ {\partial}_1{\sigma}^{11}+{\partial}_2{\sigma}^{12}+{\partial}_3{\sigma}^{13}=0,\hspace{3mm} {\partial}_1{\sigma}^{21}+{\partial}_2{\sigma}^{22}+{\partial}_3{\sigma}^{23}=0, \hspace{3mm} {\partial}_1{\sigma}^{31}+{\partial}_2{\sigma}^{32}+{\partial}_3{\sigma}^{33}=0 \]

A direct computational approach has been provided by Eugenio Beltrami in 1892 ([4]), James Clerk Maxwell in 1870 ([22]) and Giacinto Morera in 1892 by introducing the $6$ {\it stress functions} ${\phi}_{ij}={\phi}_{ji}$ in the {\it Beltrami parametrization} with formal notations:\\ 
\[   \left(  \begin{array}{r}
{\sigma}^{11} \\
{\sigma}^{12}\\
 {\sigma}^{13}\\
{\sigma}^{22}\\
{\sigma}^{23}\\
{\sigma}^{33}
\end{array}
\right)  =
\left( \begin{array}{cccccc}
 0  & 0  & 0 &d_{33}& -2d_{23}&  d_{22} \\
 0 & -d_{33}&d_{23}& 0  & d_{13}  &  -d_{12}  \\
0  & d_{23}& -d_{22}& -d_{13} & d_{12} & 0 \\
 d_{33}&0 &-2d_{13}&  0&0 &d_{11} \\
-d_{23}&d_{13}&d_{12}&0 & -d_{11}& 0 \\
 d_{22}&  -2d_{12}&  0&d_{11}&0&0
\end{array} \right)
\left(  \begin{array}{l} 
{\Phi}_{11}  \\
{\Phi}_{12}  \\
{\Phi}_{13}  \\
{\Phi}_{22}  \\
{\Phi}_{23} \\
{\Phi}_{33}
\end{array}  \right)    \]
It is involutive with $3$ equations of class $3$, $3$ equations of class $2$ and no equation of class $1$. The $3$ CC are describing the stress equations which admit therefore a parametrization, but without any geometric framework, in particular without any possibility to imagine that the above second order operator is {\it nothing else but} the {\it formal adjoint} of the {\it Riemann operator}, namely the (linearized) Riemann tensor with $n^2(n^2-1)/2=6$ independent components when $n=3$ [27-30]. 
We may rewrite the Beltrami parametrization of the Cauchy stress equations as follows, after exchanging the third row with the fourth row and using formal notations:  \\
\[      \left(  \begin{array}{cccccc}
d_1& d_2 & d_3 &0 & 0 & 0 \\
 0 & d_1 &  0 & d_2 & d_3 & 0 \\
 0 & 0 & d_1 & 0 & d_2 & d_3 
\end{array}  \right)  
 \left(  \begin{array}{cccccc}
 0 & 0 & 0 & d_{33} & - 2d_{23} & d_{22} \\
 0 & - d_{33} & d_{23} & 0 & d_{13} & - d_{12}  \\
 0 & d_{23} & - d_{22} & - d_{13} & d_{12} & 0 \\
 d_{33}& 0 & - 2 d_{13} & 0 & 0 & d_{11}  \\
 - d_{23} & d_{13} & d_{12}& 0 & - d_{11} & 0 \\
 d_{22} & - 2 d_{12} & 0 & d_{11}& 0 & 0 
 \end{array} \right)  \equiv   0    \]
 as an identity where $0$ on the right denotes the zero operator. However, the standard implicit summation used in continuum mechanics (See [40] for more details) is, when $n=3$:  \\
 \[   {\sigma}^{ij}{\Omega}_{ij} = {\sigma}^{11}{\Omega}_{11} + 2 {\sigma}^{12}{\Omega}_{12} + 2 {\sigma}^{13}{\Omega}_{13} + {\sigma}^{22} {\Omega}_{22} + 2{\sigma}^{23}{\Omega}_{23} + {\sigma}^{33}{\Omega}_{33}    \]
because {\it the stress tensor density $\sigma$ is supposed to be symmetric} in continuum mechanics. Integrating by parts in order to construct the adjoint operator, we get the striking identification:  \\
\[       Riemann=ad(Beltrami)   \hspace{1cm} \Longleftrightarrow  \hspace{1cm}   Beltrami=ad(Riemann)  \]
between the (linearized ) Riemann tensor and the Beltrami parametrization. \\
As we already said, the brothers E. and F. Cosserat proved in 1909 that the assumption ${\sigma}^{ij}={\sigma}^{ji}$ may be too strong because it only takes into account density of forces and ignores density of couples, and the Cauchy stress equations {\it must} be replaced by the so-called {\it Cosserat couple-stress equations} ([10, 31, 34]). In any case, taking into account the factor $2$ involved by multiplying the second, third and fifth row by $2$, we get the new $6\times 6$ matrix with rank $3$:
\[   \left(  \begin{array}{cccccc}
 0 & 0 & 0 & d_{33} & - 2d_{23} & d_{22} \\
 0 & - 2d_{33} & 2d_{23} & 0 & 2d_{13} & - 2d_{12}  \\
 0 & 2d_{23} & - 2d_{22} & - 2d_{13} & 2d_{12} & 0 \\
 d_{33}& 0 & - 2 d_{13} & 0 & 0 & d_{11}  \\
 - 2d_{23} & 2d_{13} & 2d_{12}& 0 & - 2d_{11} & 0 \\
 d_{22} & - 2 d_{12} & 0 & d_{11}& 0 & 0 
 \end{array} \right)     \]
This is a symmetric matrix and the corresponding second order operator with constant coefficients is thus self-adjoint. \\
{\it Surprisingly}, the Maxwell parametrization is obtained by keeping only ${\phi}_{11}=A, {\phi}_{22}=B, {\phi}_{33}=C$ while setting ${\phi}_{12}={\phi}_{23}={\phi}_{31}=0$ and using only the columns $1 + 4 + 6$ as follows:  \\
\[   \left(  \begin{array}{r}
{\sigma}^{11} \\
{\sigma}^{12}\\
 {\sigma}^{13}\\
{\sigma}^{22}\\
{\sigma}^{23}\\
{\sigma}^{33}
\end{array}
\right)  =
\left( \begin{array}{ccc}
 0 & d_{33}& d_{22}  \\
 0  &  0  & - d_{12}  \\
 0 & - d_{13} & 0  \\
 d_{33}& 0 & d_{11}  \\
 -d_{23} & 0  & 0  \\
 {\partial}_{22}&  d_{11} & 0
\end{array} \right)
\left(  \begin{array}{l} 
A  \\
B  \\
C
\end{array}  \right)    \]
and we let the reader check the corresponding Cauchy equations.\\

\noindent
$\bullet$ n=4: It is only now that we are able to explain the relation of this striking result with Einstein equations but the reader must already understand that, if we need to revisit in such a deep way the mathematical foundations of elasticity theory, we also need to revisit in a similar way the mathematical foundations of EM and GR as in ([35-37, 39-42]). To begin with, let us notice that "{\it Einstein equations are just a way to parametrize the Cauchy stress equations} ". Starting with the well known linear map $C : S_2T^* \rightarrow S_2T^*: R_{ij} \rightarrow E_{ij}=R_{ij}- \frac{1}{2}{\omega}_{ij}tr(R)$ between symmetric covariant tensors, where $\omega$ is a metric with $det(\omega)\neq 0$ and $tr(R)={\omega}^{ij}R_{ij}$, we may introduce the linear second order operators $Ricci: \Omega \rightarrow E$ and $Einstein: \Omega \rightarrow E$ obtained by linearization over $\omega$ and we have the relation $Einstein = C \circ Ricci $ where $C$ does not depend on any conformal factor. We recall the method used in any textbook for studying gravitational waves, which "{\it surprisingly} " brings the same map $ C :\Omega \rightarrow \bar{\Omega}=\Omega -\frac{1}{2}\omega \,tr(\Omega)$ in order to introduce the key unknown composite operator ${\cal{X}}:\bar{\Omega}\rightarrow \Omega \rightarrow E$, having therefore 
$Einstein = {\cal{X}} \circ C$ (See [12] for explicit formulas). Now, Theorem 2.4 proves that the {\it Einstein} operator cannot be parametrized ((36,52]) and that each component of the Weyl tensor is a torsion element killed by the Dalembertian ([8,14,41]).   \\

We finally prove that only the use of {\it algebraic analysis}, a mixture of differential geometry (differential sequences, formal adjoint) and homological algebra (module theory, biduality, extension modules) {\it totally unknown by physicists}, is able to explain why the Einstein operator (with 6 terms) defined above is useless as it can be replaced by the Ricci operator (with 4 terms) in the search for gravitational waves equations. Indeed, using the fact that the {\it Einstein} operator is self-adjoint (with a slight abuse of language) when $\omega$ is the Minkowski metric (contrary to the {\it Ricci} operator as we shall see) and taking the respective (formal) adjoint operators, we get:  \\
\[  ad(Einstein)=ad(C) \circ ad({\cal{X}}) \Rightarrow Einstein= C \circ ad({\cal{X}})\Rightarrow ad({\cal{X}})=Ricci \Rightarrow 
{\cal{X}}= ad(Ricci)  \]
Meanwhile, the {\it Riemann} operator can be considered as an operator describing the (second order) CC for the {\it Killing} operator $\xi \in T \rightarrow {\cal{L}}(\xi) \omega=\Omega \in S_2T^*$ with standard notations where ${\cal{L}}$ is the Lie derivative. In this new framework, we no longer need to use the {\it Bianchi} operator as the first order CC for the {\it Riemann} operator and thus the $Cauchy$ operator has {\it nothing to do} with the $div$-type operator induced by the $Bianchi$ operator. Also, we notice that the {\it relative parametrization} with $div$-type {\it differential constraints} needed in order to keep only the $Dalembert$ operator in the wave equations has {\it nothing to do} with any gauge transformation but has {\it only to do} with the search for a {\it minimal parametrization}, exactly like Maxwell did in 1870 for elasticity. Our final comment at the end of the case $n=2$ being still valid, we may say:  \\
{\it These purely mathematical results question the origin and existence of gravitational waves} ([41]).\\

It remains to prove that, in this new framework, the Ricci tensor only depends on the symbol ${\hat{g}}_2\simeq T^*\subset S_2T^*\otimes T$ of the first prolongation ${\hat{R}}_2\subset J_2(T) $ of the conformal Killing system ${\hat{R}}_1\subset J_1(T)$ with symbol ${\hat{g}}_1\subset T^*\otimes T$ defined by the equations ${\omega}_{rj}{\xi}^r_i + {\omega}_{ir}{\xi}^r_j - \frac{2}{n}{\omega}_{ij}{\xi}^r_r=0$ {\it not depending on any conformal factor}. In the next general commutative diagram covering both situations while taking into account that the PD equations of both the classical and conformal Killing systems are homogeneous, the {\it Spencer map} $\delta$ is induced by $-D$ and all the sequences are exact but perhaps the left column with $\delta$-cohomology $H^2(g_1)\neq 0$ at ${\wedge}^2T^*\otimes g_1$ (See later on and [27-30,48] for the definition of the {\it Spencer operator} $D$):  \\
\[  \begin{array}{rcccccccl}
   &  0 & & 0 & & 0 &  &  &   \\
   & \downarrow & & \downarrow & & \downarrow & & &  \\
0\rightarrow & g_3 & \rightarrow &  S_3T^*\otimes T & \rightarrow & S_2T^*\otimes F_0& \rightarrow & F_1 & \rightarrow 0  \\
   & \hspace{2mm}\downarrow  \delta  & & \hspace{2mm}\downarrow \delta & &\hspace{2mm} \downarrow \delta & & &  \\
0\rightarrow& T^*\otimes g_2&\rightarrow &T^*\otimes S_2T^*\otimes T & \rightarrow &T^*\otimes T^*\otimes F_0 &\rightarrow & 0 &  \\
   &\hspace{2mm} \downarrow \delta &  &\hspace{2mm} \downarrow \delta & &\hspace{2mm}\downarrow \delta &  &  &   \\
0\rightarrow & {\wedge}^2T^*\otimes g_1 & \rightarrow & \underline{{\wedge}^2T^*\otimes T^*\otimes T} & \rightarrow & {\wedge}^2T^*\otimes F_0 & \rightarrow & 0 &  \\
   &\hspace{2mm}\downarrow \delta  &  & \hspace{2mm} \downarrow \delta  &  & \downarrow  & &  &  \\
0\rightarrow & {\wedge}^3T^*\otimes T & =  & {\wedge}^3T^*\otimes T  &\rightarrow   & 0  &  &  &   \\
    &  \downarrow  &  &  \downarrow  &  &  &  &  &  \\
    &  0  &   & 0  & &  &  &  &
\end{array}  \]
We obtain at once from a snake-type chase the isomorphism $F_1 \simeq H^2(g_1)$ and provide a new simple proof of the following important result (Compare to [29, 30, 36, 37, 41] and the Remark below): \\

\noindent
{\bf THEOREM 2.13}: Introducing the $\delta$-cohomologies $H^2(g_1)$ at ${\wedge}^2T^*\otimes g_1$ and $H^2({\hat{g}}_1)$ at ${\wedge}^2T^*\otimes {\hat{g}}_1$ while taking into account that $g_1\subset {\hat{g}}_1$, we have the short exact sequences:   \\
\[   0 \rightarrow S_2T^* \rightarrow H^2(g_1) \rightarrow H^2({\hat{g}}_1) \rightarrow 0 \hspace{5mm} \Leftrightarrow  \hspace{5mm}
 0 \rightarrow S_2T^* \rightarrow F_1 \rightarrow {\hat{F}}_1 \rightarrow 0      \]   \\

\noindent
{\it Proof}: The first result can be deduced from a delicate unusual chase in the following commutative diagram where only the rows and the right column are short exact sequences. The first step is made by a diagonal snake-type chase for defining the left morphism and we let the reader check that it is a monomorphism. The right morphism is described by the inclusion ${\wedge}^2T^*\otimes g_1\subset {\wedge}^2T^*\otimes {\hat{g}}_1$ induced by the inclusion $g_1 \subset {\hat{g}}_1$ by showing that any element of ${\wedge}^2T^*\otimes {\hat{g}}_1$ is a sum of an element in ${\wedge}^2T^*\otimes g_1$ plus the image by $\delta$ of an element in 
$T^*\otimes {\hat{g}}_2$ for the right epimorphism (exercise).  \\
 \[ \begin{array}{rcccccl}
 & & & & & 0 & \\
  & & & & & \downarrow & \\
   & & & 0& & S_2T^* &  \\
  & & & \downarrow & &\hspace{2mm} \downarrow  \delta  &  \\
   & 0 &\rightarrow &T^*\otimes {\hat{g}}_2  & \rightarrow & T^*\otimes T^* & \rightarrow 0  \\
 & \downarrow & &\hspace{2mm} \downarrow \delta  & &  \hspace{2mm}\downarrow  \delta  &  \\
 0 \rightarrow & {\wedge}^2T^*\otimes g_1  & {\rightarrow} & {\wedge}^2T^*\otimes {\hat{g}}_1 & \rightarrow & {\wedge}^2T^* & \rightarrow 0  \\
   &\hspace{2mm}\downarrow \delta   & &  \hspace{2mm}\downarrow \delta  & & \downarrow     &   \\
 0 \rightarrow &{\wedge}^3T^*\otimes T &  =  & {\wedge}^3T^*\otimes T &{\rightarrow} & 0 &     \\
  & \downarrow &  & \downarrow & & &  \\
  & 0 & & 0 & & &  \\
  & & &  & & &     
   \end{array}  \]
Using the previous diagram, we obtain the isomorphisms $F_1\simeq H^2(g_1)$ and ${\hat{F}}_1\simeq H^2({\hat{g}}_1)$. We have thus the splitting sequence $ 0 \rightarrow S_2T^* \rightarrow F_1 \rightarrow {\hat{F}}_1 \rightarrow 0  $ providing a totally unusual interpretation of the successive Ricci, Riemann and Weyl tensors. It follows that $dim({\hat{F}}_1)=n(n+1)(n+2)(n-3)/12$ and the $Weyl$-type operator is of order $3$ when $n=3$ but of order $2$ for $n\geq 4$. Similar results could be obtained for the $Bianchi$-type operator. \\
\hspace*{12cm}   Q.E.D   \\

 \noindent
{\bf REMARK 2.14}: Using the contraction $T^*\otimes T \rightarrow {\wedge}^0T^* \rightarrow 0$, namely ${\xi}^ k_i \rightarrow {\xi}^r_r$,  in order to describe the cokernel of the left vertical monomorphism, we obtain the following commutative and exact diagram which is only depending on the first order jets of $T$:  \\
 \[  \begin{array}{rcccccl}
   & 0   &   &  0 &   &  &  \\
   &  \downarrow  &  & \downarrow  &  &  &   \\
 0  \rightarrow &  g_1     &  \rightarrow & T^* \otimes T & \rightarrow & F_0  & \rightarrow 0  \\     
   &  \downarrow  &  &  \downarrow  &  &  \downarrow &   \\
 0  \rightarrow & {\hat{g}}_1     &  \rightarrow & T^* \otimes T & \rightarrow & {\hat{F}}_0  & \rightarrow 0  \\ 
   &  &  &  \downarrow &  &  \downarrow &  \\
   &   &  &  0  &  &  0  &  
   \end{array}  \]
 Prolonging twice to the jets of order $3$ of $T$, we obtain the commutative and exact diagram:  \\
 \[ \begin{array}{rcccccl}
 & & & 0 & & 0 & \\
  & & &  \downarrow  & & \downarrow & \\
   & 0 & \rightarrow & S_2T^*& = & S_2T^* & \rightarrow 0 \\
  & \downarrow & & \downarrow & &\downarrow   &  \\
   0 \rightarrow &S_3T^*\otimes T &\rightarrow & S_2T^*\otimes  F_0 & \rightarrow & F_1& \rightarrow 0  \\
 & \parallel & & \downarrow & & \downarrow  &  \\
 0 \rightarrow & S_3T^*\otimes T  & {\rightarrow} & S_2T^*\otimes {\hat{F}}_0 & \rightarrow &{\hat{F}}_1 & \rightarrow 0  \\
   &\downarrow   & & \downarrow  & & \downarrow     &   \\
    & 0 & & 0 & & 0 &  
   \end{array}  \]
 providing the same short exact sequence as in the Theorem but without any possibility to establish a link between $S_2T^*$ and a $1$-form with value in the bundle ${\hat{g}}_2$ of elations.  \\

 \newpage

\noindent
{\bf EXAMPLE 2.15}: {\it Electromagnetism}.   \\ 
Passing now to electromagnetism and the original Gauge Theory (GT) which is still, up to now, the only known way to establish a link between EM and group theory, the idea is to introduce first the {\it nonlinear gauge sequence}:
\[  \begin{array}{ccccc}
X\times G & \longrightarrow & T^*\otimes {\cal{G}} &\stackrel{MC}{ \longrightarrow} & {\wedge}^2T^*\otimes {\cal{G}}  \\
a                & \longrightarrow  &    a^{-1}da=A         &    \longrightarrow & dA-[A,A]=F
\end{array}   \]
where $X$ is a manifold, $G$ is a Lie group with identity $e$ {\it not acting on} $X$, $a:X \rightarrow G$ a map identified with a section of the trivial bundle $X\times G$ over $X$ and $a^{-1}da=A$ is the pull-back over $X$ by the tangent mapping $T(a)$ of a basis of left invariant $1$-forms on $G$. Also, $[A(\xi),A(\eta)]\in {\cal{G}}, \forall \xi,\eta \in T$ by introducing the bracket on the Lie algebra ${\cal{G}}=T_e(G)$ and the pull-back of the Maurer-Cartan (MC) equations on $G$ is the so-called {\it curvature} $2$-form with value in ${\cal{G}}$. Choosing $a$ close to $e$, that is $a(x)=e+t\lambda(x)+...$ with $t\ll 1$ and linearizing as usual, we obtain the linear operator $d:{\wedge}^0T^*\otimes {\cal{G}}\rightarrow {\wedge}^1T^*\otimes {\cal{G}}:({\lambda}^{\tau}(x))\rightarrow ({\partial}_i{\lambda}^{\tau}(x))$ leading to the {\it linear gauge sequence}:\\ 
\[  {\wedge}^0T^*\otimes {\cal{G}}\stackrel{d}{\longrightarrow} {\wedge}^1T^*\otimes {\cal{G}} \stackrel{d}{\longrightarrow} {\wedge}^2T^*\otimes{\cal{G}} \stackrel{d}{\longrightarrow} ... \stackrel{d}{\longrightarrow} {\wedge}^nT^*\otimes {\cal{G}}\longrightarrow  0   \]
which is the tensor product by ${\cal{G}}$ of the Poincar\'{e} sequence for the exterior derivative $d$. In 1954, at the birth of GT, the above notations were coming from electromagnetism with EM {\it potential} $A\in T^*$ and EM {\it field} $dA=F\in {\wedge}^2T^*$ in the relativistic Maxwell theory. Accordingly, $G=U(1)$ (unit circle in the complex plane)$\longrightarrow dim ({\cal{G}})=1$ was the {\it only possibility} existing before 1970 to get a {\it pure} $1$-form $A$ (EM potential) and a {\it pure} $2$-form $F$ (EM field) when $G$ is abelian. However, this result is {\it not coherent at all} with elasticity theory as we saw and, {\it a fortiori}, with the analytical mechanics of rigid bodies where the Lagrangian is a quadratic expression of such 1-forms when $n=3$ and $G=S0(3)$ (Compare to [26] and [3]).  \\
 
 Before going ahead, let us prove that there may be mainly two types of differential sequences, the {\it Janet sequence} introduced by M. Janet in 1920 ([15]), having to do with the tools we have studied, and a different sequence called {\it Spencer sequence} introduced by D. C. Spencer in 1970 ([48] for the linear framework, [17] for the linear framework) with totally different operators. For this, if $E$ is a vector bundle over the base $X$, we shall introduce the $q$-jet bundle $J_q(E)$ over $X$ with (local) sections ${\xi}_q: (x)\rightarrow ({\xi}^k(x), {\xi}^k_i(x), {\xi}^k_{ij}(x), ... )$ transforming like the (local) sections $j_q(\xi):(x) \rightarrow ({\xi}^k(x), {\partial}_i{\xi}^k(x), {\partial}_{ij}{\xi}^k_(x), ...) $.  When $T=T(X)$ is the tangent bundle of $X$, the {\it Spencer operator} $D:J_{q+1}(E) \rightarrow T^*\otimes J_q(T)$ and its extension $D:{\wedge}^rT^*\otimes J_{q+1}(E) \rightarrow {\wedge}^{r+1}T^*\otimes  J_q(E)$ defined by $D(\alpha \otimes {\xi}_{q+1})=d\alpha \otimes {\xi}_q + (-1)^r\alpha \wedge D{\xi}_{q+1}$, allow to compare these sections by considering the differences $({\partial}_i{\xi}^k(x)-{\xi}^k_i(x), {\partial}_i{\xi}^k_j(x) - {\xi}^k_{ij}(x), ...)$ and so on. When $\omega$ is a nondegenerate metric with Christoffel symbols $\gamma$ and Levi-Civita isomorphism $j_1(\omega)\simeq (\omega, \gamma)$, we consider the second order involutive system $R_2\subset J_2(T)$ defined by considering the first order Killing system ${\cal{L}}(\xi)\omega=0$, adding its first prolongation ${\cal{L}}(\xi)\gamma=0$ and using ${\xi}_2$ instead of $j_2(\xi)$. Looking for the first order generating {\it compatibility conditions} (CC) ${\cal{D}}_1$ of the corresponding second order operator operator ${\cal{D}}$ just described, we may then look for the generating CC ${\cal{D}}_2$ of ${\cal{D}}_1$ and so on, exactly like in the differential sequence made successively by the {\it Killing}, {\it Riemann}, {\it Bianchi}, ... operators. We may proceed similarly for the injective operator $T \stackrel{j_2}{\longrightarrow} C_0(T)=J_2(T)$, finding successively $C_0(T) \stackrel{D_1}{\longrightarrow}C_1(T)$ and $C_1(T) \stackrel{D_2}{\longrightarrow}C_2(T)$ induced by $D$. When $n=2$ and $\omega$ is the $Euclide$ metric, we have a Lie group of isometries with the $3$ infinitesimal generators $\{{\partial}_1, {\partial}_2, x^1{\partial}_2 - x^2{\partial}_1\}$. If we now consider the Weyl group defined by ${\cal{L}}(\xi)\omega = A\omega$ with $A=cst$ and ${\cal{L}}(\xi)\gamma=0$, we have to add the only dilatation $x^1{\partial}_1 + x^2 {\partial}_2$. Collecting the results and exhibiting the induced kernel upper differential sequence, we get the following commutative {\it fundamental diagram} where the upper down arrows are monomorphisms while the lower down arrows are epimorphisms ${\Phi}_0, {\Phi}_1, {\Phi}_2$:  \\

 \[  \begin{array}{rccccccccccccr}
  & 0& \longrightarrow& \tilde{\Theta} &\stackrel{j_2}{\longrightarrow}& 4 &\stackrel{D_1}{\longrightarrow}& 8 &\stackrel{D_2}{\longrightarrow} &  4 &\longrightarrow  0 &  \\

   & 0& \longrightarrow& \Theta &\stackrel{j_2}{\longrightarrow}& 3 &\stackrel{D_1}{\longrightarrow}& 6 &\stackrel{D_2}{\longrightarrow} &  3 &\longrightarrow  0 & \hspace{3mm}Spencer  \\
  &&&&& \downarrow & & \downarrow & & \downarrow & &    & \\
   & 0 & \longrightarrow &   2   & \stackrel{j_2}{\longrightarrow} &  12  & \stackrel{D_1}{\longrightarrow} &   16  &\stackrel{D_2}{\longrightarrow} &  6   &   \longrightarrow 0 &\\
   & & & \parallel && \hspace{5mm}\downarrow {\Phi}_0 & &\hspace{5mm} \downarrow {\Phi}_1 & & \hspace{5mm}\downarrow {\Phi}_2 &  &\\
   0 \longrightarrow & \Theta &\longrightarrow &   2  & \stackrel{\cal{D}}{\longrightarrow} &  9  & \stackrel{{\cal{D}}_1}{\longrightarrow} & 10 & \stackrel{{\cal{D}}_2}{\longrightarrow} &   3  & \longrightarrow  0 & \hspace{7mm} Janet \\
   
    0 \longrightarrow & \tilde{\Theta} &\longrightarrow &   2  & \stackrel{{\cal{D}}}{\longrightarrow} &  8  & \stackrel{{\cal{D}}_1}{\longrightarrow} & 8  & \stackrel{{\cal{D}}_2}{\longrightarrow} &   2  & \longrightarrow  0 & 
   \end{array}     \]
   
\vspace{5mm}   
\noindent
It follows that "{\it Spencer and Janet play at see-saw} ", the dimension of each {\it Janet bundle} being decreased by the same amount as the dimension of the corresponding {\it Spencer bundle} is increased, this number being the number of additional parameters multiplied by $dim({\wedge}^rT^*)$. \\

More generally, whenever $R_q\subseteq J_q(E)$ is an involutive system of order $q$ on $E$, we may define the {\it Janet bundles} $F_r$ for $r=0,1,...,n$ by the short exact sequences:  \\
\[0 \rightarrow {\wedge}^rT^*\otimes R_q+\delta ({\wedge}^{r-1}T^*\otimes S_{q+1}T^*\otimes E)\rightarrow {\wedge}^rT^*\otimes J_q(E) \rightarrow F_r \rightarrow 0  \]
We may pick up a section of $F_r$, lift it up to a section of ${\wedge}^rT^*\otimes J_q(E)$ that we may lift up to a section of ${\wedge}^rT^*\otimes J_{q+1}(E)$ and apply $D$ in order to get a section of ${\wedge}^{r+1}T^*\otimes J_q(E) $ that we may project onto a section of $F_{r+1}$ in order to construct an operator ${\cal{D}}_{r+1}:F_r\rightarrow F_{r+1}$ generating the CC of ${\cal{D}}_r$ in the canonical {\it linear Janet sequence}:  \\
\[  0 \longrightarrow  \Theta \longrightarrow E \stackrel{\cal{D}}{\longrightarrow} F_0 \stackrel{{\cal{D}}_1}{\longrightarrow}F_1 \stackrel{{\cal{D}}_2}{\longrightarrow} ... \stackrel{{\cal{D}}_n}{\longrightarrow} F_n \longrightarrow 0   \]
If we have two involutive systems $R_q \subset {\hat{R}}_q \subset J_q(E)$, {\it the Janet sequence for} $R_q$ {\it projects onto the Janet sequence for} ${\hat{R}}_q$ and we may define inductively {\it canonical epimorphisms} $F_r \rightarrow {\hat{F}}_r \rightarrow 0$ for 
$r=0, 1,...,n$ by comparing the previous sequences for $R_q$ and ${\hat{R}}_q$, as we already saw.  \\ 
A similar procedure can also be obtained if we define the Spencer bundles $C_r$ for $r=0,1,...,n$ by the short exact sequences:  \\
\[ 0 \rightarrow  \delta ({\wedge}^{r-1}T^*\otimes g_{q+1} )  \rightarrow  {\wedge}^rT^*\otimes R_q  \rightarrow  C_r  \rightarrow 0 \]
We may pick up a section of $C_r$, lift it to a section of ${\wedge}^rT^*\otimes R_q$, lift it up to a section of ${\wedge}^rT^*\otimes R_{q+1}$ and apply $D$ in order to construct a section of ${\wedge}^{r+1}\otimes R_q$ that we may project to $C_{r+1}$ in order to construct an operator $D_{r+1}:C_r \rightarrow C_{r+1}$ generating the CC of $D_r$ in the canonical {\it linear Spencer sequence} which is {\it another completely different resolution} of the set $\Theta$ of (formal) solutions of $R_q$:  \\
\[    0 \longrightarrow \Theta \stackrel{j_q}{\longrightarrow} C_0 \stackrel{D_1}{\longrightarrow} C_1 \stackrel{D_2}{\longrightarrow} C_2 \stackrel{D_3}{\longrightarrow} ... \stackrel{D_n}{\longrightarrow} C_n\longrightarrow 0  \]
However, if we have two systems as above, {\it the Spencer sequence for} $R_q$ {\it is now contained into the Spencer sequence for} 
${\hat{R}}_q$ and we may construct inductively {\it canonical monomorphisms} $0\rightarrow C_r \rightarrow {\hat{C}}_r$ for $r=0,1,...,n$ by comparing the previous sequences for $R_q$ and ${\hat{R}}_q$.   \\

When dealing with applications, we have set $E=T$ and considered systems of finite type Lie equations determined by Lie groups of transformations. 
{\it In this specific case}, it can be proved that the Janet and Spencer sequences are formally exact, both with their respective adjoint sequences ([30,34,40,41,42]), namely $ad({\cal{D}}_r)$ generates the CC of $ad({\cal{D}}_{r+1})$ while $ad(D_r)$ generates the CC of $ad(D_{r+1})$. 
We have obtained in particular $C_r={\wedge}^rT^*\otimes R_q \subset {\wedge}^rT^*\otimes {\hat{R}}_q ={\hat{C}}_r$ when comparing the classical and conformal Killing systems, but {\it these bundles have never been used in physics}. Therefore, instead of the classical Killing system $R_2\subset J_2(T)$ defined by $\Omega \equiv {\cal{L}}(\xi)\omega=0$ {\it and} $\Gamma\equiv {\cal{L}}(\xi)\gamma=0$ or the conformal Killing system ${\hat{R}}_2\subset J_2(T)$ defined by $\Omega\equiv {\cal{L}}(\xi)\omega=A(x)\omega$ and ${\Gamma} \equiv {\cal{L}}(\xi)\gamma= ({\delta}^k_iA_j(x) +{\delta} ^k_j A_i(x) -{\omega}_{ij}{\omega}^{ks}A_s(x)) \in S_2T^*\otimes T$, we may introduce the {\it intermediate differential system} ${\tilde{R}}_2 \subset J_2(T)$ defined by ${\cal{L}}(\xi)\omega=A\omega$ with $A=cst$ and $\Gamma \equiv {\cal{L}}(\xi)\gamma=0 $, for the {\it Weyl group} obtained by adding the only dilatation with infinitesimal generator $x^i{\partial}_i$ to the Poincar\'e group, exactly like we already did when $n=2$. We have $R_1\subset {\tilde{R}}_1={\hat{R}}_1$ but the strict inclusions $R_2 \subset {\tilde{R}}_2 \subset {\hat{R}}_2$ and we discover {\it exactly} the group scheme used through this paper, both with the need to {\it shift by one step to the left} the physical interpretation of the various differential sequences used. Indeed, as ${\hat{g}}_2\simeq T^*$ because ${\xi}^r_{ri}(x)=n A_i(x)$, the first Spencer operator ${\hat{R}}_2\stackrel{D_1}{\longrightarrow} T^*\otimes {\hat{R}}_2$ is induced by the usual Spencer operator ${\hat{R}}_3 \stackrel{D}{\longrightarrow} T^*\otimes {\hat{R}}_2:(0,0,{\xi}^r_{rj},{\xi}^r_{rij}=0) \rightarrow (0,{\partial}_i0-{\xi}^r_{ri}, {\partial}_i{\xi}^r_{rj}- 0)$ and thus projects by cokernel onto the induced operator $T^* \rightarrow T^*\otimes T^*$. Composing with $\delta$, it projects therefore onto $T^*\stackrel{d}{\rightarrow} {\wedge}^2T^*:A \rightarrow dA=F$ as in EM and so on by using the fact that $D_1$ 
{\it and} $d$ {\it are both involutive}, or the composition of epimorphisms:  \\
\[ {\hat{C}}_r \rightarrow {\hat{C}}_r/{\tilde{C}}_r\simeq {\wedge}^rT^*\otimes ({\hat{R}}_2/{\tilde{R}}_2) \simeq {\wedge}^rT^*\otimes {\hat{g}}_2\simeq {\wedge}^rT^*\otimes T^*\stackrel{\delta}{\longrightarrow}{\wedge}^{r+1}T^* \]
The main result we have obtained is thus to be able to increase the order and dimension of the underlying jet bundles and groups, proving therefore that any $1$-form with value in the second order jets ${\hat{g}}_2$ ({\it elations}) of the conformal Killing system (conformal group) can be decomposed uniquely into the direct sum $(R,F)$ where $R$ is a section of the {\it Ricci bundle} $S_2T^*$ and the EM field $F$ is a section of ${\wedge}^2T^*$ (Compare to [53]).  \\

Lippmann got the Nobel prize in 1908 for the discovery of color photography. Only one year
later, in 1909, the brothers E. and F. Cosserat wrote their ``{\it Th\'eorie des corps d\'eformables}" ([10]) and it is in this book 
that the previous analogies are quoted for the first time. Between 1895 and 1910, the two brothers published together a series of
Notes in the ``Comptes Rendus de l'Acad\'emie des Sciences de Paris'' and long Notes in famous textbooks or treatises on the mathematical 
foundations of elasticity theory [49]. In particular, they proved that one can exhibit all the concepts and formulas to be
found in {\it elasticity theory} (deformation/strain, compatibility conditions, stress, stress equations, constitutive relations, ...) just by 
knowing the group of rigid motions of ordinary 3-dimensional space with $3$ translations and $3$ rotations. \\

It is rather astonishing that {\it all the formulas} that can be found in the book written by E. and F. Cosserat in 1909 are nothing else but the
formal adjoint of the Spencer operator for the Killing equations. More precisely, a section ${\xi}_2$ of the first prolongation 
$R_2\subset J_2(T)$ of the system $R_1\subset J_1(T)$ of {\it Killing equations} is a section of the 2-jet bundle $J_2(T)$ of the
tangent bundle $T=T(X)$, namely a set of functions ${\xi}^k(x), {\xi}^k_i(x),{\xi}^k_{ij}(x)$, transforming like the derivatives
${\xi}^k(x), {\partial}_i{\xi}^k(x), {\partial}_{ij}{\xi}^k(x)$ of a vector field $\xi$ but also satisfying the linear equations:  \\
\[   {\omega}_{rj}{\xi}^r_i(x)+{\omega}_{ir}{\xi}^r_j(x)+{\xi}^r(x){\partial}_r{\omega}_{ij}=0,\hspace{1cm} {\xi}^k_{ij}(x)=0 \]
where $\omega$ is the euclidean metric. Multiplying by test fuctions $\sigma$ and $ \mu $ respectively the zero and first order components of 
the image $D{\xi}_2$ of the corresponding Spencer operator $D$, then integrating by part while moving up and down the
dumb indices by means of the metric, we successively obtain:  \\
\[  \begin{array}{rcl}  
 {\sigma}^i_k({\partial}_i{\xi}^k-{\xi}^k_i)+
{\mu}^{j,i}_k({\partial}_i{\xi}^k_j-{\xi}^k_{ij}) &
= & {\sigma}^{ir}{\partial}_r{\xi}_i-{\sigma}^{ij}{\xi}_{i,j}+
{\mu}^{ij,r}{\partial}_r {\xi}_{i<j}\\
  & = & -[({\partial}_r{\sigma}^{ir}){\xi}_i+({\partial}_r{\mu}^{ij,r}+
{\sigma}^{ij}-{\sigma}^{ji}){\xi}_{i<j}]+divergence \nonumber
\end{array}  \]
\[   \Rightarrow   {\partial}_r{\sigma}^{ir}= f^i, \\hspace{5mm}  {\partial}_r{\mu}^{ij,r}   +{\sigma}^{ij} - {\sigma}^{ji}= m^{ij}   \]                                                          
with evident notations for the Einstein summations involved (Compare to [10],p 137 and 167).\\ 

Keeping in mind that, in space-time, there are 4 translations $({\xi}^k)$ and 6 rotations $({\xi}^k_i)$ (3 space rotations + 3 Lorentz transformations), 
we recover all the $4+6+1+4=15$ variations that can be found in the engineering calculus leading to finite element computations (MODULEF library for example). In addition, we have proved in many books ([29,30,32,42]) and papers ([35,37,39]) that the conformal group of space-time is the biggest group of invariance of the Minkowski constitutive laws of EM in vacuum while both sets of Maxwell equations are invariant by any diffeomorphism. In particular, considering the space-time dilatation $x^i\rightarrow a\,x^i$ for $i=1,2,3,4$ with infinitesimal generator $x^i{\partial}_i$, a transformation which has no intuitive meaning, and gauging the connected compoment $[0,+\infty[$ of the identity with the distinguished identity $1$, that is to say transforming the group parameters into functions, just explains why {\it there must be a zero lower bound in the measure of absolute temperature, 
both with a distinguished value and  invariance under} $T\longrightarrow 1/T$.\\

 This result clarifies the Helmholtz analogy within jet theory. Indeed, if $T$ is identified with the inverse of a first jet of dilatation, then $T$ 
{\it behaves like the derivative of a function without being such a proper derivative}, and we find again exactly the definition of a jet coordinate. 
{\it Such a result should lead in the future to revisit the foundations of thermostatics/thermodynamics} ([39]). \\
The additional 4 transformations, called {\it elations,} are highly nonlinear and we understand
that, contrary to E. and F. Cosserat who succeeded in dealing with the linear transformations, H. Weyl did not succeed in relating
electromagnetism with the second order jets of the conformal group in ([51]), {\it though the idea was a genious one}, simply 
because he could not use in 1920 a mathematical tool created in 1970 ([17,48]) but only effective in 1983 ([28-30,37,39]).\\ 
The reader may now understand that such a geometric unification was indeed the dream of the brothers E. and F. Cosserat who refer many times explicitly to the work of Mach and Lippmann ([10],p 147,211]). More precisely, using now the {\it conformal Killing equations}, we have:  \\
\[ {\omega}_{rj}{\xi}^r_i(x)+{\omega}_{ir}{\xi}^r_j(x)+{\xi}^r(x){\partial}_r{\omega}_{ij}=A(x){\omega}_{ij},\hspace{5mm} 
{\xi}^k_{ij}(x)={\delta}^k_iA_j(x)+{\delta}^k_jA_i(x)-{\omega}_{ij}{\omega}^{kr}A_r(x) \]
where $A(x)$ is an arbitrary function and $A_i(x)dx^i$ is an arbitrary 1-form, we get ${\xi}^r_{ri}(x)=nA_i(x)$ and ${\xi}^k_{ijr}(x)=0$ for
$n\geq 3$ [27-29]. Accordingly, the zero, first and second order components ({\it field}) of the image $D{\xi}_3$ of the Spencer operator $D$ are:  \\
\[ {\partial}_i{\xi}^k-{\xi}^k_i, \hspace{1cm} {\partial}_i{\xi}^k_j-{\xi}^k_{ij},\hspace{1cm} {\partial}_r{\xi}^k_{ij}-{\xi}^k_{ijr}=
{\partial}_r{\xi}^k_{ij} \,\,\,  \Rightarrow \,\,\,  {\partial}_i{\xi}^r_{rj} - {\xi}^r_{rij}={\partial}_i{\xi}^r_{rj}       \]
and we can recover ${\epsilon}_{ij}=\frac{1}{2}[({\partial}_i{\xi}_j-{\xi}_{i,j})+({\partial}_j{\xi}_i-{\xi}_{j,i})]$.
Identifying the speed with a (gauged) Lorentz rotation, that is to say setting ${\partial}_4{\xi}^k-{\xi}^k_4=0$ as a constraint ([11]), we can 
therefore measure both ${\partial}_4{\xi}^k_4-{\xi}^k_{44}=(1/c^2){\gamma}^k-{\omega}^{kr}A_r$ for
$k=1,2,3$ (care to the sign !) and ${\partial}_i{\xi}^r_r-{\xi}^r_{ri}=-n((1/T){\partial}_iT+A_i)$, thus 
$(1/T)\vec{\nabla}T+(1/c^2)\vec{\gamma}$ by substraction, where $\vec{\gamma}$ is the acceleration, and thus $(1/T)\vec{\nabla}T$ in first 
approximation ([11],p.922). Also, the formula ${\partial}_i{\xi}^r_{rj}-{\partial}_j{\xi}^r_{ri}=n({\partial}_iA_j-
{\partial}_jA_i)=nF_{ij}$ exactly describes the results of [51] by means of the Spencer operator and explains why the EM field is on equal footing with
deformation and gradient of temperature, contrary to its status in gauge theory.\\
{\it Roughly speaking, E. and F. Cosserat were only using the zero and first order components of the image of the Spencer operator while H. Weyl 
was only using the first and second order components} (See [34] for more comments and [28-30,35,39] for a nonlinear version).\\

\vspace{2cm}

\noindent
{\bf 4) CONCLUSION}  \\

Recapitulating all the results obtained, we may say:  \\
$\bullet$ "{\it Beyond the mirror} " of the classical approach to apparently well known and established theories, there is a totally new interpretation of these theories and their couplings by means of the Spencer sequence for the conformal Killing operator.  \\
$\bullet$ The purely mathematical results of Section $3$ perfectly agree with the origin and existence of elastic and electromagnetic waves but question the origin and existence of gravitational waves because the parametrization of the {\it Cauchy} operator can be simply done by the adjoint of the {\it Ricci} operator without any reference to the {\it Einstein} operator.   \\
$\bullet$ They also prove that the concept of "{\it field} " in a physical theory must not be related with the concept of " {\it curvature} " because it is a $1$-form with value in a Lie algebroid and {\it not} a $2$ form with value in a Lie algebra.   \\
$\bullet$ They finally prove that gravitation and electromagnetism have a common conformal origin. In particular, electromagnetism has only to do with the conformal group of space-time and not with $U(1)$ as it is still believed today in Gauge Theory.  \\

\newpage

\noindent
{\bf REFERENCES}   \\

\noindent
[1] Adler, F.W.: \"{U}ber die Mach-Lippmannsche Analogie zum zweiten Hauptsatz, Anna. Phys. Chemie, 22 (1907) 578-594.  \\
\noindent
[2] Airy, G.B.:  On the Strains in the Interior of Beams, Phil. Trans. Roy. Soc.London, 153 (1863) 49-80.  \\  
\noindent
[3] Arnold, V.: M\'{e}thodes Math\'{e}matiques de la M\'{e}canique Classique, Appendice 2 (G\'{e}od\'{e}siques des m\'{e}triques invariantes \`{a} gauche sur des groupes de Lie et hydrodynamique des fluides parfaits), MIR, Moscow (1974,1976). \\
\noindent
[4] Beltrami, E.: Osservazioni sulla Nota Precedente, Atti Reale Accad. Naz. Lincei Rend., 5 (1892) 141-142.  \\
\noindent
[5] Bjork, J.E. (1993) Analytic D-Modules and Applications, Kluwer (1993).  \\ 
\noindent
[6] Bourbaki, N.: Alg\`{e}bre, Ch. 10, Alg\`{e}bre Homologique, Masson, Paris (1980). \\
\noindent
[7] de Broglie, L.: Thermodynamique de la Particule isol\'{e}e, Gauthiers-Villars, Pris (1964).  \\
\noindent
[8] Choquet-Bruhat, Y.: Introduction to General Relativity, Black Holes and Cosmology, Oxford University Press (2015).  \\
\noindent
[9] Chyzak, F., Quadrat, A., Robertz, D.: {\sc OreModules}: A symbolic package for the study of multidimensional linear systems,
Springer, Lecture Notes in Control and Inform. Sci., 352 (2007) 233-264.\\
http://wwwb.math.rwth-aachen.de/OreModules  \\
\noindent
[10] Cosserat, E., \& Cosserat, F.: Th\'{e}orie des Corps D\'{e}formables, Hermann, Paris, (1909).\\
\noindent
[11] Eckart, C.: The Thermodynamics of irreversible Processses, Phys. Rev., 58 (1940) 919-924.  \\
\noindent
[12] Foster, J., Nightingale, J.D.: A Short Course in General relativity, Longman (1979).  \\
\noindent
[13] Hu,S.-T.: Introduction to Homological Algebra, Holden-Day (1968).  \\
\noindent
[14] Hughston, L.P., Tod, K.P.: An Introduction to General Relativity, London Math. Soc. Students Texts 5, Cambridge University Press 
(1990). \\
\noindent
[15] Janet, M.: Sur les Syst\`{e}mes aux D\'{e}riv\'{e}es Partielles, Journal de Math., 8 (1920) 65-151. \\
\noindent 
[16] Kashiwara, M.: Algebraic Study of Systems of Partial Differential Equations, M\'{e}moires de la Soci\'{e}t\'{e} 
Math\'{e}matique de France, 63 (1995) (Transl. from Japanese of his 1970 MasterÕs Thesis).  \\
\noindent
[17] Kumpera, A., \& Spencer, D.C.: Lie Equations, Ann. Math. Studies 73, Princeton University Press, Princeton (1972).\\
\noindent
[18] Lippmann, G.: Extension du Principe de S. Carnot \`{a} la Th\'{e}orie des P\'{e}nom\`{e}nes \'{e}lectriques, C. R. Acad/ Sc. Paris, 82 (1876) 1425-1428.  \\
\noindent
[19] Lippmann, G.: \"{U}ber die Analogie zwischen Absoluter Temperatur un Elektrischem Potential, Ann. Phys. Chem., 23 (1907) 994-996. \\ 
\noindent
[20] Mach, E.: Die Geschichte und die Wurzel des Satzes von der Erhaltung der Arbeit, p 54, Prag: Calve (1872).  \\
\noindent
[21]ÊMach, E.: Prinzipien der W\"{a}rmelehre, 2, Aufl., p 330, Leipzig: J.A. Barth (1900).  \\
\noindent
[22] Maxwell, J.C.: On Reciprocal Figures, Frames and Diagrams of Forces, Trans. Roy. Soc. Ediinburgh, 26 (1870) 1-40.  \\
\noindent
[23] Northcott, D.G.: An Introduction to Homological Algebra, Cambridge university Press (1966).  \\
\noindent
[24] Oberst, U.: Multidimensional Constant Linear Systems, Acta Appl. Math., 20, 1-175 (1990).  \\ 
\noindent
[25] Ougarov, V.: Th\'{e}orie de la Relativit\'{e} Restreinte, MIR, Moscow, 1969, (french translation, 1979).\\
\noindent
[26] Poincar\'{e}, H.: Sur une Forme Nouvelle des Equations de la M\'{e}canique, C. R. Acad\'{e}mie des Sciences Paris, 132 (7) (1901) 369-371.  \\
\noindent
[27] Pommaret, J.-F.: Systems of Partial Differential Equations and Lie Pseudogroups, Gordon and Breach, New York (1978); Russian translation: MIR, Moscow,(1983).\\
\noindent
[28] Pommaret, J.-F.: Differential Galois Theory, Gordon and Breach, New York (1983).\\
\noindent
[29] Pommaret, J.-F.: Lie Pseudogroups and Mechanics, Gordon and Breach, New York (1988).\\
\noindent
[30] Pommaret, J.-F.: Partial Differential Equations and Group Theory, Kluwer (1994).\\
http://dx.doi.org/10.1007/978-94-017-2539-2    \\
\noindent
[31] Pommaret, J.-F.: Fran\c{c}ois Cosserat and the Secret of the Mathematical Theory of Elasticity, Annales des Ponts et Chauss\'ees, 82 (1997) 59-66 (Translation by D.H. Delphenich).  \\
\noindent
[32] Pommaret, J.-F.: Partial Differential Control Theory, Kluwer, Dordrecht (2001).\\
\noindent
[33] Pommaret, J.-F.: Algebraic Analysis of Control Systems Defined by Partial Differential Equations, in "Advanced Topics in Control Systems Theory", Springer, Lecture Notes in Control and Information Sciences 311 (2005) Chapter 5, pp. 155-223.\\
\noindent
[34] Pommaret, J.-F.: Parametrization of Cosserat Equations, Acta Mechanica, 215 (2010) 43-55.\\
http://dx.doi.org/10.1007/s00707-010-0292-y  \\
\noindent
[35] Pommaret, J.-F.: Spencer Operator and Applications: From Continuum Mechanics to Mathematical Physics, in "Continuum Mechanics-Progress in Fundamentals and Engineering Applications", Dr. Yong Gan (Ed.), ISBN: 978-953-51-0447--6, InTech (2012) Available from: \\
http://dx.doi.org/10.5772/35607   \\
\noindent
[36] Pommaret, J.-F.: The Mathematical Foundations of General Relativity Revisited, Journal of Modern Physics, 4 (2013) 223-239. \\
 http://dx.doi.org/10.4236/jmp.2013.48A022   \\
  \noindent
[37] Pommaret, J.-F.: The Mathematical Foundations of Gauge Theory Revisited, Journal of Modern Physics, 5 (2014) 157-170.  \\
http://dx.doi.org/10.4236/jmp.2014.55026  \\
 \noindent
[38] Pommaret, J.-F.: Relative Parametrization of Linear Multidimensional Systems, Multidim. Syst. Sign. Process., 26 (2015) 405-437.  \\
DOI 10.1007/s11045-013-0265-0   \\
\noindent
[39] Pommaret,J.-F.:From Thermodynamics to Gauge Theory: the Virial Theorem Revisited, pp. 1-46 in "Gauge Theories and Differential geometry,", NOVA Science Publisher (2015).  \\
\noindent
[40] Pommaret, J.-F.: Airy, Beltrami, Maxwell, Einstein and Lanczos Potentials revisited, Journal of Modern Physics, 7 (2016) 699-728. \\
\noindent
http://dx.doi.org/10.4236/jmp.2016.77068   \\
\noindent
[41] Pommaret, J.-F.:Why Gravitational Waves Cannot Exist, Journal of Modern Physics, 8 (2017) 2122-2158.  \\
https://doi.org/104236/jmp.2017.813130    \\
\noindent
[42] Pommaret, J.-F.: Deformation Theory of Algebraic and Geometric Structures, Lambert Academic Publisher (LAP), Saarbrucken, Germany (2016). A short summary can be found in "Topics in Invariant Theory ", S\'{e}minaire P. Dubreil/M.-P. Malliavin, Springer 
Lecture Notes in Mathematics, 1478 (1990) 244-254.\\
http://arxiv.org/abs/1207.1964  \\
\noindent
[43] Pommaret, J.-F. and Quadrat, A.: Localization and Parametrization of Linear Multidimensional Control Systems, Systems \& Control Letters, 37 (1999) 247-260.  \\
\noindent 
[44] Quadrat, A.: An Introduction to Constructive Algebraic Analysis and its Applications, 
Les cours du CIRM, Journees Nationales de Calcul Formel, 1(2), 281-471 (2010).\\
\noindent
[45] Quadrat, A., Robertz, R.: A Constructive Study of the Module Structure of Rings of Partial Differential Operators, Acta Applicandae Mathematicae, 133 (2014) 187-234. \\
http://hal-supelec.archives-ouvertes.fr/hal-00925533   \\
\noindent
[46] Rotman, J.J.: An Introduction to Homological Algebra, Pure and Applied Mathematics, Academic Press (1979).  \\
\noindent
[47] Schneiders, J.-P.: An Introduction to D-Modules, Bull. Soc. Roy. Sci. Li\`{e}ge, 63, 223-295 (1994).  \\
\noindent
[48] Spencer, D.C.: Overdetermined Systems of Partial Differential Equations, Bull. Am. Math. Soc., 75 (1965) 1-114.\\
\noindent
[49] Teodorescu, P.P.: Dynamics of Linear Elastic Bodies,  Abacus Press, Tunbridge, Wells (1975) 
(Editura Academiei, Bucuresti, Romania).\\
\noindent
[50] Vessiot, E.: Sur la Th\'{e}orie des Groupes Infinis, Ann. Ec. Norm. Sup., 20 (1903) 411-451.   \\ 
http://numdam.org  \\
\noindent
[51] Weyl, H.: Space, Time, Matter, Springer (1918,1958); Dover (1952). \\
\noindent
[52] Zerz, E.: Topics in Multidimensional Linear Systems Theory, Lecture Notes in Control and Information Sciences, Springer, LNCIS 256 (2000) \\
\noindent
[53] Zou, Z., Huang, P., Zang ,Y., Li, G.: Some Researches on Gauge Theories of Gravitation, Scientia Sinica, XXII, 6 (1979) 628-636.\\

\end{document}